\documentstyle[12pt]{article}
\begin{document}
\vskip .7cm
\begin{center}
{\large{\bf {$CP^{1}$ model with Hopf interaction: the quantum theory}}}
\vskip 2cm

{\bf B. Chakraborty$^{*,\dagger }$, Subir Ghosh$^{{**},\ddagger}$ and
R. P. Malik $^{*, \P}$}
\vskip 0.5cm
$^{*}$ {\it S. N. Bose National Centre for Basic Sciences,} \\
{\it Block-JD, Sector-III, Salt Lake, Calcutta-700 098, India}\\
$^{**}$ {\it  Physics and Applied Mathematics Unit,}\\
{\it Indian Statistical Institute, 203 B. T. Road, Calcutta-700 035, 
India}\\

\vskip 2.5cm

\end{center}

\noindent
{\bf Abstract} \\

The $CP^1$ model with Hopf interaction 
is quantised following the Batalin-Tyutin (BT) prescription.  In this
scheme, extra BT fields are introduced which allow for the existence of 
only commuting first-class constraints. Explicit expression for the  quantum 
correction to the expectation value of the energy 
density and angular momentum in the physical sector of this model is derived. 
The result shows, in the particular operator ordering that 
we have chosen to work with, that the quantum effect has a divergent 
contribution of $ {\cal O} (\hbar^2)$ in the energy expectation value. 
But, interestingly the Hopf term, though topological in nature, can have 
a  finite ${\cal O} (\hbar)$ contribution to
energy density in the homotopically nontrivial topological sector.
The angular momentum operator, however, is found to have no
quantum correction, indicating the absence of any fractional
spin even at this quantum level. Finally, the extended Lagrangian
incorporating the BT auxiliary fields is computed in the conventional 
framework of BRST formalism exploiting Faddeev-Popov technique of
path integral method.

\vskip 1cm


\baselineskip=16pt

\vskip 2cm

\footnotetext{$^{\dagger}$ E-mail address: biswajit@boson.bose.res.in}
\footnotetext{$^{\ddagger}$ E-mail address : sghosh@isical.ac.in}
\footnotetext{$^{\P}$ E-mail address: malik@boson.bose.res.in}

\newpage

\noindent
{\bf 1 Introduction}\\

\noindent
Constrained dynamical systems that one encounters usually in physical theories
may or may not involve first class constraint(s)(FCC) (in the Dirac
classification scheme \cite{dir}) which are weakly in involution. For 
example, the free Maxwell theory has two FCCs, whereas $O(3)$ Non-Linear
Sigma Model (NLSM) has none. Systems with FCC(s)
correspond to gauge theories, where the FCC(s) by
themselves play the role of generator(s) of gauge transformations. In 
this sense, NLSM is not a gauge theory since it involves only
a pair of second class constraints(SCC) which are non-commuting in the
classical Poisson bracket sense. However, there is nothing very sacred 
in this distinction between gauge invariant and gauge non-invariant theories, 
since one can elevate the latter to an equivalent former type by 
introducing extra 
degrees of freedom. A well known example is the equivalence between the
$O(3)$ NLSM and $CP^1$ model, where the latter is a $U(1)$ gauge theory 
which has an enlarged phase space. 
The usual prescription for quantising theories with SCCs
is to implement strongly these SCCs by using Dirac brackets (DB). These
DBs provide the canonical symplectic structure
for the theory. Thereafter the procedure for quantisation is quite
straightforward in ``principle''. One has to just elevate these DBs 
into quantum commutators ($\{\; , \;\}_{DB}
\rightarrow {1\over {i\hbar}}[\; , \;]$), where
the dynamical variables now correspond to operators. For gauge 
theories, one gets some additional subsidiary conditions by demanding that
physical Hilbert space is a gauge invariant subspace of the total 
Hilbert space and therefore corresponds to the kernel of the FCCs, i.e., the 
states belonging to physical Hilbert states are annihilated by these FCCs.

In contrast to the above mentioned Dirac quantisation, these gauge 
systems can also be quantised in the reduced phase space scheme, where the 
already existing FCC's are rendered SCC's by using a fresh set of DB's. The 
symplectic structure therefore undergoes further modification. In this scheme, 
the system loses all the gauge degrees of freedom and one can isolate only the 
physical degrees of freedom and work with them. However, this symplectic
structure, associated to this reduced phase space may
become field dependent so that its
elevation to quantum algebras is beset with operator ordering 
ambiguities. These complications also arise in models with nonlinearities.
For example, in NLSM or its equivalent $CP^1$ model (to be studied here),
the symplectic structure given by the basic DBs are field dependent. For 
$CP^1$ model, the DBs appropriate for the reduced phase space will be even
more complicated.

One can bypass these problems by following the Batalin, Fradkin and 
Vilkovisky \cite{bfv} scheme where the phase space is enlarged still
further, rather than reduced, by introducing some additional fields in
such a manner that even the existing SCCs are now elevated to FCCs.
This is quite akin to the idea mentioned at the beginning of this Section.
 The advantage with this scheme is that one can just work with basic PBs where
there is no operator ordering problem.  A particular construction by
Batalin and Tyutin (BT) \cite{bt} in this regard is very appealing since 
this scheme renders the FCC algebra in the extended phase space completely
Abelian. A number of applications, specifically in NLSM and $CP^N$ 
models \cite{bbg}, highlights the complexities in nonlinear models. BT have 
further provided a systematic way of constructing first-class operators that
commute with the (converted) FCC's. This formalism has been used by 
Hong, Kim and Park \cite{kor} who have constructed the first-class (improved)
version of the phase space variables and have shown that the extended
infinite series Hamiltonian \cite{bbg} can be summed to a compact form 
at least for the $O(3)$ and $CP^1$ models. Furthermore,
the advantage of this scheme is that the
$CP^1$ constraint maintains its form in terms of the improved variables.
Besides, this being true also for the Hamiltonian, the existence of
solitonic configurations \cite{sol} are naturally ensured in the 
extended phase space by virtue of the fact
that Bogomol'nyi inequality will also retain its original form.

The present paper deals with the BT quantisation of the $CP^1$ model 
with Hopf interaction. In a recent paper \cite{kor1}, the $CP^1$ model 
(without the Hopf term) has been studied using the BT prescription. However, 
the analysis
is not correct since the original FCC present in the gauge invariant
$CP^1$ model has been overlooked. Obviously, the equivalence between 
$O(3)$ NLSM and $CP^1$ model is lost even at the level of degree of freedom 
count (if the FCC is not taken into account).
This constraint, along with other ones, has a direct bearing on the
construction of the BRST charge $Q_{B}$ which defines the physical 
states ($ Q_{B} |phys> = 0 $) in the extended Hilbert space of states. 
Indeed, the presence of the Hopf term apart from  contributing to more 
complexities in a technical sense (as the constraints and their algebras 
are modified), gives rise to very interesting physical consequences.
For example, the $CP^1$ model with its solitonic
solutions, has major implications in the realm of condensed matter 
physics and the $O(3)$ NLSM describes anti-ferromagnetic
systems having a linear dispersion relations. On the other hand,
its {\it non-relativistic} version describes a ferromagnetic system
having a quadratic dispersion relation \cite{fradkin}. The solitons
in this ferromagnetic system may correspond to the skyrmions in a
quantum Hall system. It has been shown in \cite{gov} that the Hopf
term alters the spin algebra drastically. Also the
inclusion of the (topological) Hopf term in the NLSM (in its usual
relativistic version) has been shown in  \cite{wz}
 to impart fractional spin to the soliton. In a quantum analysis, using
the path integral method, it was shown in  \cite{wz} that the system 
acquires a non-trivial phase upon a spatial rotation by $2\pi$.
Since the Hopf term contribution, i.e., the fractional spin, appeared in
$O(\hbar^0)$, it seems that this result should be derivable 
in a classical (Dirac) analysis.  
But a canonical Dirac analysis \cite{bkw} revealed fractional 
spin in the above model only after the model was altered using a
certain identity which is not a constraint in the Dirac sense. 
This analysis was essentially carried out at a classical level, as the
structure of the Dirac brackets were too complicated to lend
themselves to be elevated into quantum commutators; the expressions
were marred by the operator ordering ambiguities. This has
been pointed out in  \cite{cm}. It was
further shown \cite{cm} that fractional spin was not induced by the Hopf
term alone if the above mentioned identity was not used.
The same conclusion can be drawn even at the level of collective 
coordinate quantisation (as we show in the next Section). 
It is therefore desirable to study the model at quantum level and 
investigate whether any fractional spin of order
$ {\cal O} (\hbar)$ emerges or not (as a classical treatment did not 
reveal any fractional spin \cite{cm}). In either case, therefore,
the result will be different from \cite{bkw} as follows from a dimensional
analysis. BT quantisation is adopted here to 
avoid of the above mentioned operator ordering ambiguities 
appearing in the Dirac brackets.
As a first step towards this goal, in the present paper, we study the 
quantum correction to the energy of this model following the approach of 
\cite{no}. 
One does not expect the Hopf term (being a topological term) to contribute to
the energy-momentum tensor. Here we get a surprising result that the Hopf term
may contribute non-trivially to the energy density in the homotopically
nontrivial topological sector at the quantum level. We then take up the
case of angular momentum to find that there is no quantum correction. 
Another reason for considering $CP^1$ model, rather than NLSM, 
is that the Hopf term is local in terms of $CP^1$ variables \cite{wuz} 
where no gauge-fixing condition is required {\it a priori}.
The model is thus a $U(1)$ gauge theory and amenable to gauge independent
Dirac quantization. As we have mentioned earlier that the structure of the
Dirac brackets in the presence of SSCs is less complicated in the Dirac scheme
than the corresponding gauge-fixed reduced phase space scheme. Besides, in the
BT scheme, even the existing SCCs are going to be elevated to FCC, so that the
symplectic structure is field independent and is canonical as it is obtained
from the simple PB.

The paper is organised as follows: In Section II the collective 
coordinate quantisation of the $CP^1$ model with the Hopf term is discussed. 
It has been shown
that the Hopf term does not have any effect on the energy or spin of the 
soliton. Section III is devoted to the classical constraint analysis. Section 
IV deals with the Batalin-Tyutin extension where the FCCs, First
Class (FC)  variables and the FC Hamiltonian are constructed in the extended 
phase space. The important results regarding the presence and absence of Hopf 
term induced quantum correction  
to the energy density and angular momentum respectively,
are derived in Section V. Conventional BRST quantisation is 
outlined in Section VI. The internal consistency of the results is 
also checked by re-deriving the 
action in the unitary gauge. The paper ends with a conclusion and future 
perspectives in Section VII.\\

\noindent
{\bf 2 Collective co-ordinate quantisation}\\

\noindent
In this Section, we are going to provide a brief review of $CP^1$ fields
and their relationship with the fields of NLSM, apart from setting up
our conventions. Furthermore,  we are going to show
that in the $CP^1$ model, fractional spin is not induced by the
(topological) Hopf term at the level of collective coordinate 
quantisation,
 unless the model is {\it altered} by  using an identity,
(which is not a Hamiltonian constraint), as has been done in \cite{bkw}.

The $CP^1$ manifold is given by the set of all non-zero complex doublets
$Z  = \left ( \begin{array}{c} z_1  \\ z_2 \\ \end{array} \right ) $
satisfying the normalization condition
$$
\begin{array}{lcl}
Z^{\dagger} Z = |z_{1}|^2 + |z_{2}|^2 = 1,
\end{array} \eqno(2.1)
$$
and the identification $ Z \sim e^{i \theta} \; Z$ where $e^{i\theta} 
\in U(1)$ is any unimodular number in the complex plane. Since eqn. (2.1)
represents $S^3$, which is homeomorphic to $SU(2)$ group manifold, 
$CP^1$ space can be identified with the coset space $ \frac{SU(2)} {U(1)}$. 
Alternatively,
$SU(2)$ can be identified with a $U(1)$ prinicipal bundle over the base 
$CP^1$.  Clearly, one can make a local (in the 
neighbourhood $z_{1} \neq 0)$ gauge 
choice
$$
\begin{array}{lcl}
z_{1} -  z_{1}^*  = 0,
\end{array} \eqno(2.2)
$$
so that (2.1) is reduced to an equation of $S^2$ enabling one to
finally identify $CP^1$ with $S^2$. This gauge will be also discussed at the 
end (in Section VI).

Associated with this $U(1)$ bundle, there is a natural $U(1)$ connection
 one form \cite{cm,bc}
$$
\begin{array}{lcl}
A = - i Z^\dagger \; d \; Z,
\end{array} \eqno(2.3)
$$
and can be identified with the Dirac magnetic monopole connection
one-form.  Now the $CP^1$ model in $(2 + 1)$ dimensions is 
\footnote{ We adopt the notations in which the flat
Minkowski metric is:  $ g_{\mu\nu}$ = diag $ ( +1, -1, -1)$ and
the totally antisymmetric Levi-Civita tensor satisfies:
$ \varepsilon^{012} = \varepsilon_{012} = + 1,\;
\varepsilon_{0ij} = \varepsilon_{ij}, \; \varepsilon^{12}
= \varepsilon_{12} = + 1$.
Here, and in what follows, the Greek indices $\mu, \nu, \rho....
= 0, 1, 2$ and the Latin indices $ i, j, k......= 1, 2$ correspond
to space-time and space directions respectively on the space-time 
manifold.
The summation convention on the subscript $\alpha (= 1, 2)$ is 
occasionally supressed.}
$$
\begin{array}{lcl}
L_{0} = {\displaystyle \int}\; d^2 x\; \bigl [
(D_{\mu} Z)^\dagger (D^\mu Z) - \lambda (Z^\dagger Z - 
1 \bigr ],
\end{array} \eqno(2.4a)
$$
where the covariant derivative operator $D_{\mu}$ is given by
$$
\begin{array}{lcl}
D_{\mu} = \partial_{\mu} - i\; A_{\mu},
\end{array} \eqno(2.4b)
$$
and $A_{\mu} = -i Z^\dagger \partial_{\mu} Z$ is nothing but the 
pull-back of the connection one-form (2.3) onto the spacetime
and $\lambda$ is a Lagrange multiplier enforcing the constraint
(2.1). The form (2.4a) can be further simplified to
$$
\begin{array}{lcl}
L_{0} = {\displaystyle \int} \; d^2 x\;
\Bigl [\; |\partial_{\mu} Z|^2 - |Z^\dagger \partial_{\mu} Z|^2
- \lambda\; (|Z|^2 - 1) \;\Bigr ].
\end{array} \eqno(2.5)
$$
Formally, this $CP^1$ model is same as NLSM given by
$$
\begin{array}{lcl}
L_{0}^\prime  = {\displaystyle \int} \; d^2 x\;
\Bigl [ \;\frac{1}{4}\; \partial_{\mu} n^T\;\partial^\mu n
- \lambda\; (n^T n - 1) \Bigr ].
\end{array} \eqno(2.6)
$$
where $ n =
\left (\begin{array}{c}
n_{1}\\
n_{2}\\ n_{3}\\
\end{array} \right )$ represents a set of three real 
scalar fields subjected to the constraint
$$
\begin{array}{lcl}
n^T n \equiv n_{a} n_{a} = n_{1}^2 + n_{2}^2 + n_{3}^2 = 1.
\end{array} \eqno(2.7)
$$
The equivalence between $CP^1$ model and NLSM can be trivially 
established \cite{cm} (see also Rajaraman in \cite{sol}) by noting 
the fact that these $n_{a}$'s can be obtained from
the $CP^1$ fields by using the Hopf map
$$
\begin{array}{lcl}
n_{a} = Z^\dagger \;\sigma_{a}\; Z,
\end{array} \eqno(2.8)
$$
where $\sigma$'s are the Pauli matrices. Although equivalent, $CP^1$ 
model is a $U(1)$ gauge theory but NLSM is not. Correspondingly, $CP^1$ fields 
$Z$ transform nontrivially ($ Z \rightarrow e^{i\phi} Z$) under $U(1)$ gauge 
transformation, but the fields $ n_{a}$ of NLSM  are invariant under $U(1)$
transformations as is clear from (2.8).

In order to obtain a finite energy static solution, it is necessary for the
fields $n_{a}$ to tend to constant configuration asymptotically. With this,
the two-dimensional plane ${\cal D}$ gets effectively compactified to $S^2$
and the configuration space $C$, which is nothing but the set of all maps
$ f : S^2 \rightarrow S^2$ (field manifold) splits into a disjoint union
of path connected spaces as
$$
\begin{array}{lcl}
\Pi_{0} (C) = \Pi_{2} (S^2) = Z.
\end{array} \eqno(2.9)
$$
Hence there exists solitons or skyrmions in this model, characterized by
the set of integers $ Q \in Z$ given by
$$
\begin{array}{lcl}
Q = {\displaystyle \int}\; d^2 x\; j^{0},
\end{array} \eqno(2.10)
$$
where $j^{0}$ is the time component of the identically conserved 
($ \partial_{\mu} j^\mu = 0$) topological current ($ j^\mu$) given by
$$
\begin{array}{lcl}
j^\mu = \frac{1}{8\pi}\;
\varepsilon^{\mu\nu\lambda} \varepsilon_{abc}
n_{a} \partial_{\nu} n_{b} \partial_{\lambda} n_{c}.
\end{array} \eqno(2.11)
$$
Note that the conservation of $j^\mu$ holds irrespective of any equation
of motion. $Q$, referred to as the soliton number, labels the disconnected 
pieces of the configuration space $C$. This topological current (2.11) can
also be expressed as the curl of the $U(1)$ gauge field (2.3) and thus,
in terms of the $CP^1$ fields, as
$$
\begin{array}{lcl}
j^\mu &=& \frac{1}{2\pi}\;
\varepsilon^{\mu\nu\lambda} 
\partial_{\nu} A_{\lambda}, \nonumber\\
&=& - \frac{i}{2\pi}\; \varepsilon^{\mu\nu\lambda}
\partial_{\nu} Z^\dagger \partial_{\lambda} Z,
\end{array} \eqno(2.12)
$$
In any soliton number sector, the fundamental group of $C$ is nontrivial
since
$$
\begin{array}{lcl}
\Pi_{1} (C) = \Pi_{3} (S^2) = Z.
\end{array} \eqno(2.13)
$$
This implies that the loops based at any point in the configuration space
fall into separate homotopy classes labeled by another integer $H$. This
integer can be given a representation by the so-called Hopf action
\cite{forte}
$$
\begin{array}{lcl}
S_{H} = {\displaystyle \int}\; d^3 x\;
j^\mu A_{\mu}.
\end{array} \eqno(2.14)
$$
Although it is formally similar to the Chern-Simon (CS) action, there is
subtle difference in the sense that in Hopf action, unlike in CS action, the
gauge field is not an independent variable in configuration space and is
determined in terms of $j^\mu$ using certain gauge condition to render it
a generic non-local current-current interaction. Thus, in Hopf case, we do
not have an enlarged phase space unlike the CS case. 
However, the advantage of $CP^1$ formulation
is that the Hopf action (2.14) can be expressed in a gauge invariant manner,
which is {\it not} non-local. The hermitian form of Hopf action is
$$
\begin{array}{lcl}
S_{H} = - \frac{1}{4 \pi}\;{\displaystyle \int}\; d^3 x\;
\varepsilon^{\mu\nu\lambda} \bigl (
Z^\dagger \partial_{\mu} Z - \partial_{\mu} Z^\dagger Z \bigr )
\partial_{\nu} Z^\dagger \; \partial_{\lambda} Z.
\end{array} \eqno(2.15)
$$

In the topological sector $ Q = 1$, the $n_{a}$ fields can be 
taken to have the profile
$$
\begin{array}{lcl}
n = \left ( 
\begin{array}{c}
\hat r sin ~ g(r)\\
 cos ~ g(r) \\
\end{array} \right )
\end{array} \eqno(2.16)
$$
where $ \hat r$ being the unit vector in the $ n_{1}, n_{2}$ plane and
$ g (r)$ satisfies
$$
\begin{array}{lcl}
g (0) = 0, \;\;\;\qquad \;\;\;g(\infty) = \pi.
\end{array} \eqno(2.17)
$$
The corresponding profile for the $Z$-fields can be trivially obtained 
by inverting (2.8) in  a particular gauge. For example, in the gauge (2.2),
the $CP^1$ fields have the following profile
$$
\begin{array}{lcl}
Z = \left ( 
\begin{array}{c}
cos \frac{g(r)} {2}\\
 sin \frac{g(r)} {2}\;
e^{i (\phi + \alpha (t))} \\
\end{array} \right  )
\end{array} \eqno(2.18)
$$
with $ \phi = tan^{-1} (\frac{n_{2}} {n_{1}}) $ being the polar angle 
in the $n_{1}, n_{2}$ plane. As any configuration, obtained by making an 
$SO(2)$ rotation by an angle $\alpha$ in (2.16) is 
energetically degenerate to (2.16), we 
introduce a corresponding $U(1)$ factor $ e^{i \alpha (t)}$ in (2.18) where
the phase $\alpha (t) $ is the collective coordinate in this case.

Substituting (2.18) in (2.5), one gets after a straightforward calculation
$$
\begin{array}{lcl}
L_{0} = \frac{\pi}{2}\; \lambda\; \dot \alpha^2 - N,
\end{array} \eqno(2.19)
$$
where
$$
\begin{array}{lcl}
N &=& \frac{\pi}{2}\; {\displaystyle \int}_{0}^{\infty}\; dr\;
r \; \Bigl [ (g^{\prime} (r))^2 + \frac{1}{r^2}\; sin^2 g(r) \Bigr ],
\nonumber\\
\lambda &=& {\displaystyle \int}_{0}^{\infty}\; dr\; r\;
sin^2 g(r).
\end{array} \eqno(2.20)
$$
This is expected to be the same as the corresponding expression one gets
using NLSM \cite{bkw}. However for the same profile of the $Z$ fields as 
in (2.18), the Hopf term \cite{cm}
$$
\begin{array}{lcl}
L_{H} = \theta \; {\displaystyle \int}\; d^2 x\; 
\varepsilon^{\mu\nu\lambda}\;
\Bigl [ Z^\dagger \partial_\mu Z \partial_{\nu} Z^\dagger 
\partial_{\lambda} Z
+ \partial_{\mu} Z^\dagger Z \partial_{\lambda} Z^\dagger
\partial_{\nu} Z \Bigr ],
\end{array} \eqno(2.21)
$$
obtained from (2.15) with the inclusion of Hopf parameter $\theta$, vanishes.
This is true without performing the space-time integration.
But if the Hopf term is {\it altered} a la' \cite{bkw} using the identity
$$
\begin{array}{lcl}
{\displaystyle \int}\; d^2 x\; A_{0} (x) j_{0} (x) = -
{\displaystyle \int}\; d^2 x\; A_{i} (x) j_{i} (x),
\end{array} \eqno(2.22a)
$$
or written in terms of $CP^1$ variables as
$$
\begin{array}{lcl}
{\displaystyle \int}\; d^2 x \;Z^\dagger \dot Z
\varepsilon_{ij} (\partial_{i} Z)^\dagger (\partial_{j} Z)
= {\displaystyle \int} \;d^2 x\; \varepsilon_{ij} Z^\dagger
(\partial_{i} Z) \; \bigl [ (\partial_{j} Z)^\dagger \dot Z -
\dot Z^\dagger \partial_{j} Z \bigr ],
\end{array} \eqno(2.22b)
$$
to get
$$
\begin{array}{lcl}
\tilde L_{H} = \frac{\theta } {\pi}\;
{\displaystyle \int}\; d^2 x\; \varepsilon^{i \nu\lambda}\;
\partial_{\nu} Z^\dagger \partial_{\lambda} Z \;
Z^\dagger \partial_{i} Z,
\end{array} \eqno(2.23)
$$
then for the same profile (2.18) of the $CP^1$ field, one gets a 
non-vanishing $\tilde L_{H}$ as
$$
\begin{array}{lcl}
\tilde L_{H} = \theta\; \dot \alpha.
\end{array} \eqno(2.24)
$$
Note that the identity (2.22) which is valid in the radiation gauge, is 
not a Hamiltonian constraint as it involves time derivative. Thus the 
dynamical consequences
of $CP^1$ model coupled to $L_{H}$ or $\tilde L_{H}$ can be quite 
different.  It was shown in \cite{cm} that $L_{H}$, coupled to  $CP^1$, 
does not induce
any fractional spin at the classical level, whereas $\tilde L_{H}$ 
induces fractional spin at the classical level. As we shall see now
again following \cite{bkw} that the same
result follows even at the level of collective coordinate quantisation.

In order to compute fractional spin for the model
$$
\begin{array}{lcl}
 L =  L_{0} + L_{H},
\end{array} \eqno(2.25)
$$
let us first note that the Hopf term $L_{H}$ being a topological term, 
does not contribute to the symmetric expression of energy-momentum tensor
($ T_{\mu\nu} \sim\;\frac{ \delta S} {\delta g^{\mu\nu}}$). This tensor 
is given by
$$
\begin{array}{lcl}
T_{\mu\nu} = (D_{\mu} Z)^\dagger (D_\nu Z ) + (D_\nu Z)^\dagger (D_\mu 
Z) - g_{\mu\nu} (D_\rho Z)^\dagger (D_\rho Z).
\end{array} \eqno(2.26)
$$
The angular momentum $ J = \int d^2 x \; \varepsilon_{ij} x_{i} 
T_{0j}$, corresponding to the profile (2.18) of the $Z$-field, is given as
$$
\begin{array}{lcl}
J = - \pi\; \lambda \; \dot \alpha.
\end{array} \eqno(2.27)
$$
As $L_{H} = 0$, corresponding to this profile, the
Lagrangian (2.25) reduces by using (2.19) to
$$
\begin{array}{lcl}
L = L_{0} = \frac{\pi}{2}\; \lambda\; \dot \alpha^2 - N.
\end{array} \eqno(2.28)
$$
The canonically conjugate momenta is then the angular momentum
(up to a sign)
$$
\begin{array}{lcl}
p = \frac{\partial L}{\partial \dot \alpha}
= \pi\;\lambda\; \dot \alpha = - J.
\end{array} \eqno(2.29)
$$
The Legendre transformed Hamiltonian, using eqn. (2.28) gives
$$
\begin{array}{lcl}
H = p \dot \alpha - L = N + \frac{1} {2 \pi \lambda}\; J^2.
\end{array} \eqno(2.30)
$$
This can be identified with the Hamiltonian of a
rigid rotor with moment of inertia equal to $\pi \lambda$.

Clearly, the energy eigenfunctions $e^{i m \alpha}$ are also the 
eigenfunctions of $J$. The corresponding eigen values are $- m$, namely;
$$
\begin{array}{lcl}
J \; e^{im\alpha} = - p\; e^{im\alpha} = i \frac{\partial} {\partial 
\alpha}\;
e^{i m \alpha} = - m\; e^{i m \alpha}.
\end{array} \eqno(2.31)
$$
Single valuedness in $\alpha$ space restricts $m $ to be an integer ( 
i.e., $ m \in Z$). Thus, the system does not exhibit fractional spin at the 
level of collective coordinate quantisation. On the otherhand, if the $CP^1$ 
model is coupled to the {\it altered} Hopf term $\tilde L_{H}$, i.e., if we 
consider the model
$$
\begin{array}{lcl}
L^\prime = L_{0} + \tilde L_{H},
\end{array} \eqno(2.32)
$$
instead of (2.25), then for the profile (2.18) of the $Z$ field, eqn. 
(2.32), on using eqn. (2.24), reduces to
$$
\begin{array}{lcl}
L^\prime = \frac{\pi} {2}\; \lambda \; \dot \alpha^2
+ \theta \dot \alpha - N.
\end{array} \eqno(2.33)
$$
This is the same Lagrangian considered in \cite{bkw} where the existence 
of fractional spin was shown at the level of collective coordinate 
quantisation.
Note the presence of the $\theta$ dependent Hopf term here. This 
corroborates our \cite{cm} earlier observation that fractional spin can not be 
obtained unless the model is {\it altered}. But in a gauge independent Dirac 
quantisation one can not use the identity (2.22) which is valid only in the 
radiation gauge. On the other hand,
Wilczek and Zee \cite{wz} have shown in the path integral framework 
the existence of the fractional spin. In order to obtain any contribution of 
the Hopf term, related to fractional spin,
in a canonical framework, one has to go for the full quantisation
of this model. This motivates us for our present study of this model in 
the BT formalism.\\

\noindent
{\bf 3 Constraint analysis }\\

\noindent
The classical action ($S$) of the $CP^1$ model
with the Hopf interaction is obtained by adding (2.5) and (2.21) as
$$
\begin{array}{lcl}
S = {\displaystyle \int} d^3 x {\cal L}_{cl}
&=& {\displaystyle \int} d^3 x
\Bigl [ \partial_{\mu} Z^{\dagger} \partial^\mu Z
- (Z^{\dagger} \partial_\mu Z)
(\partial^\mu Z^{\dagger} \; Z) - \lambda
(Z^{\dagger}\; Z - 1) \nonumber\\
&+& \theta \varepsilon^{\mu\nu\rho}
\bigl (Z^{\dagger} \partial_{\mu} Z
\partial_{\nu} Z^{\dagger} \partial_{\rho} Z
+ \partial_{\mu} Z^{\dagger} Z \partial_{\rho}
Z^{\dagger} \partial_{\nu} Z \bigr ) \Bigr ].
\end{array}\eqno(3.1)
$$
The canonical analysis yields
the following expressions for the canonically conjugate momenta
$$
\begin{array}{lcl}
\Pi_{\alpha} &=& \frac{\partial 
{\cal L}_{cl}} {\partial \dot z_{\alpha}}
= \dot z_{\alpha}^{*} - z_{\alpha}^{*} (\dot Z^{\dagger} Z) + \theta 
M^{*}_{\alpha},
\nonumber\\
\Pi_{\alpha}^* &=& \frac{\partial {\cal L}_{cl}} {\partial \dot 
z_{\alpha}^{*}}
= \dot z_{\alpha} - z_{\alpha} (Z^\dagger \dot Z) + \theta M_{\alpha}
\end{array} \eqno(3.2a)
$$
where the expressions for $M_{\alpha}^{*}$ and $M_{\alpha}$ are:
$$
\begin{array}{lcl}
M_{\alpha}^* &=& \varepsilon_{ij} \bigl [
z_{\alpha}^* (\partial_{i} Z^\dagger \partial_{j} Z) - \partial_{j} 
z_{\alpha}^*
(\partial_{i} Z^\dagger Z) + \partial_{j} z_{\alpha}^* (Z^\dagger
 \partial_{i} Z) 
\bigr ],
\nonumber\\
M_{\alpha} &=& \varepsilon_{ij} \bigl [
z_{\alpha} (\partial_{j} Z^\dagger \partial_{i} Z) - \partial_{j} z_{\alpha}
(Z^\dagger \partial_{i} Z) + \partial_{j} z_{\alpha} (\partial_{i} Z^\dagger Z) 
\bigr ].
\end{array} \eqno(3.2b)
$$
The Legendre transformation leads to the derivation of the
classical Hamiltonian density
$$
\begin{array}{lcl}
{\cal H}_{cl} &=& \Pi_\alpha \dot z_\alpha 
+ \Pi^*_\alpha \dot z^*_\alpha - {\cal L}_{cl}, 
\nonumber\\
&=& | \dot Z |^2 - |\dot Z^\dagger Z |^2 + | \partial_{i} Z |^2
- | Z^\dagger \partial_{i} Z |^2 + \lambda ( |Z|^2 - 1 ).
\end{array} \eqno(3.3)
$$
In the above Hamiltonian density, the contribution of the Hopf term
apparently disappears since it consists of only linear time derivative 
terms.
However, $\theta$-dependent terms reappear when the phase space 
variables
are introduced to express the above Hamiltonian density as given below
$$
\begin{array}{lcl}
{\cal H}_{cl} &=& | \Pi - \theta M^* |^2 + |\partial_{i} Z|^2
- | Z^* \partial_{i} Z |^2 + \lambda (|Z|^2 - 1), \nonumber\\
&\equiv& |P|^2 + |\partial_{i} Z|^2 - |Z^\dagger \partial_{i} Z|^2
+ \lambda (|Z|^2 - 1),
\end{array} \eqno(3.4)
$$
where $ P_\alpha = (\Pi_\alpha - \theta M^*_\alpha)$ and its conjugate are
nothing but the conjugate momenta in the absence of Hopf term.
The composite variables $M's$ in (3.2b) obey the following useful identities:
$$
\begin{array}{lcl}
z_\alpha  M^*_\alpha + z^*_\alpha M_\alpha &=& 0, \nonumber\\
\dot z_\alpha M^*_\alpha + \dot z^*_\alpha M_\alpha &=& {\cal L}_{H}.
\end{array}\eqno(3.5)
$$
>From the primary constraint $T_{0} = \Pi_{\lambda} \approx 0$, the
rest of the constraints $T_i$ are derived by demanding  time
persistence of the constraints themselves as:
$$
\begin{array}{lcl}
\{ \Pi_{\lambda},   H_{cl} \} = - ( |Z|^2 - 1 ) \equiv - T_{1},
\end{array}\eqno(3.6)
$$
$$
\begin{array}{lcl}
\{ T_{1},  H_{cl} \} = z_\alpha P_\alpha 
+ z^*_\alpha P^*_\alpha \equiv - T_{2},
\end{array}\eqno(3.7)
$$
$$
\begin{array}{lcl}
\{ T_{2},  H_{cl} \} = - 4 \theta \varepsilon_{ij}
(\partial_{j} Z^\dagger \partial_{i} Z) T_{3} - 2 \lambda T_{1} + 2 T_{4},
\end{array}\eqno(3.8)
$$
$$
\begin{array}{lcl}
\{ T_{3},  H_{cl} \} = - 4 i \theta \varepsilon_{ij}
(\partial_{i} Z^\dagger \partial_{j} Z) T_{2} - i (\partial_{i} \partial_{i} 
Z^\dagger Z
- Z^\dagger \partial_{i} \partial_{i} Z) T_{1}.
\end{array}\eqno(3.9)
$$
where the expressions for $T_{3}$ and $T_{4}$ are:
$$
\begin{array}{lcl}
T_{4} &=& - \lambda + |P|^2 + 2 |\partial Z^\dagger Z|^2 
- |\partial_{i} Z|^2
+ 4 \theta \varepsilon_{ij} (\partial_{i} Z^\dagger Z) 
(P_\alpha \partial_{j} z_\alpha
- P^*_\alpha \partial_{j} z^*_\alpha ), \nonumber\\
T_{3} &=& i \bigl ( z_\alpha P_\alpha - z^*_\alpha P^*_\alpha \bigr ).
\end{array}
$$
Equations (3.6--3.9) show that the constraint algebra is closed. Before
proceeding to the Dirac classification of the constraints, we note that
$T_{0}$ and $T_{4}$ constitute an SCC pair. These
constraints are implemented strongly (in the Dirac sense) in ${\cal 
H}_{cl}$, without making any changes in the Poisson brackets of the remaining 
fields
$ z, z^*, \Pi, \Pi^*$. This leads to
$$
\begin{array}{lcl}
{\cal H}_{cl} &=& |Z|^2 |P|^2 + (2 - |Z|^2)\; |\partial_{i} Z|^2
+ (2 |Z|^2 - 3) |Z^\dagger \partial_{i} Z|^2 \nonumber\\
&+& 4 \theta \varepsilon_{ij} \;(|Z|^2 - 1)
 (\partial_{i} Z^\dagger\; Z)
(P_\alpha \partial_{j} z_\alpha - P^*_\alpha \partial_{j} z^*_\alpha).
\end{array} \eqno(3.10)
$$
Comparing with the previous results of the $CP^1$ model \cite{bbg}, it 
can be seen that the mere replacements: $ \Pi_{\alpha} \rightarrow P_{\alpha},
\Pi_{\alpha}^* \rightarrow P_{\alpha}^* $ in the expressions for
the constraints and ${\cal H}_{cl}$ \cite{bbg}, leads to the derivation 
of the corresponding expressions for the present model (with the Hopf term).

A straightforward calculation shows that
$$
\begin{array}{lcl}
T_{2} = z_\alpha P_\alpha + z^*_\alpha P^*_\alpha 
= z_\alpha \Pi_\alpha + z^*_\alpha \Pi^*_\alpha,
\end{array}\eqno(3.11)
$$
$$
\begin{array}{lcl}
T_{3} = i \bigl (z_\alpha P_\alpha - z^*_\alpha P^*_\alpha \bigr ) 
= i \Bigl (z_\alpha \Pi_\alpha - z^*_\alpha \Pi^*_\alpha  + 2 \theta 
\varepsilon_{ij}\; \bigl ( |Z|^2 \partial_{j} Z^\dagger \partial_{i} Z +
(\partial_{i} Z^\dagger \;Z) (Z^\dagger \partial_{j} Z) \bigr ) \Bigr ).
\end{array}\eqno(3.12)
$$
Now, the constraint algebra turns out to be
$$
\begin{array}{lcl}
&& \{ T_{1} (x), T_{2} (y) \} = 2 |Z|^2 \delta (x - y), \;\;\;\qquad 
\;\;
\{ T_{1} (x), T_{3} (y) \} = 0, \nonumber\\
&& \{ T_{2} (x), T_{3} (y) \} = 8 \;\theta \;\varepsilon_{ij}\;
\Bigl [ |Z|^2 \partial_{i} Z^\dagger \partial_{j} Z + (Z^\dagger 
\partial_{i} Z) (\partial_{j} Z^\dagger \;Z) \Bigr ]\; \delta(x - y).
\end{array} \eqno(3.13)
$$
Let us choose the pair $T_{1}$ and $T_{2}$ to be SCC. Since the $CP^1$ 
model has a $U(1)$ gauge invariance, the corresponding FCC, in our case, is 
derived as the following linear combination
$$
\begin{array}{lcl}
T &=& T_{3} + 4 \theta \varepsilon_{ij}
\bigl [ (\partial_{i} Z^\dagger \partial_{j} Z) + \frac{1}{|Z|^2}
(Z^\dagger \partial_{i} Z) (\partial_{j} Z^\dagger\; Z) \bigr ], \nonumber\\
&\equiv&
T_{3} + 4 \theta \varepsilon_{ij}
\bigl [ (\partial_{i} Z^\dagger \partial_{j} Z) \bigr ] T_{1}
+ 4 \theta \varepsilon_{ij} \frac{1}{|Z|^2}
(Z^\dagger \partial_{i} Z) (\partial_{j} T_{1}) T_{1}, \nonumber\\
&\equiv& T_{3} + 4 \theta \varepsilon_{ij} (\partial_{i} Z^\dagger
\partial_{j} Z) T_{1}.
\end{array} \eqno(3.14)
$$
The last term in the second line is dropped since it is quadratic in the 
constraints.

As a simple warm up exercise, one can re-derive the original action 
(3.1), starting from the Hamiltonian by exploiting the path integral expression
for the partition function ($ {\cal Z}$)

$$
\begin{array}{lcl}
{\cal Z} &=& {\displaystyle \int} {\cal D} (z, z^*, \Pi, \Pi^*)\;
\delta (T_{1})\; \delta (T_{2}) \;( \mbox {det} 
\{ T_{1}, T_{2} \} )^{\frac{1}{2}}
\delta (T)\; \delta (\chi)\; \mbox {det} \{ T, \chi \} \nonumber\\
&& \mbox {exp} \Bigl (\; i \int d^3 x \Bigl [ 
\Pi_\alpha \dot z_\alpha + \Pi^*_\alpha \dot z^*_\alpha
- ( |P|^2 + |\partial_{i} Z|^2 - |Z^\dagger 
\partial_{i} Z|^2) \Bigr ] \;\Bigr ).
\end{array} \eqno(3.15)
$$
Note that in the above expression , we do not use the more complicated 
form of the Hamiltonian in (3.10) since $ \delta (T_1)$ has been included in 
the measure. We shall follow this principle later too. In the expression 
(3.15), $\chi$ is the gauge-fixing condition corresponding to the FCC ($T$) and
the square-root factor is the Senjanovic measure which reduces to a 
c-number term on the constrained manifold. To recover the co-ordinate space 
action (3.1), we introduce the multiplier fields $\lambda_{1}, \lambda_{2}, 
\lambda_{3}$ as
$$
\begin{array}{lcl}
{\cal Z} &=& {\displaystyle \int} {\cal D} (z, z^*, \Pi, \Pi^*, \lambda_{1}
\lambda_{2}, \lambda_{3})\;
 (\mbox {det} \{ T_{1}, T_{2} \} )^{\frac{1}{2}}
\; \delta (\chi)\; \mbox {det} \{ T, \chi \} \nonumber\\
&& \mbox {exp} \Bigl ( i \int d^3 x \;\Bigl [ 
\Pi_\alpha \dot z_\alpha + \Pi^*_\alpha \dot z^*_\alpha
- ( |P|^2 + |\partial_{i} Z|^2 - |Z^\dagger \partial_{i} Z|^2) \nonumber\\
&& + \lambda_{1} (|Z|^2 - 1) + \lambda_{2} ( P_\alpha z_\alpha 
+ P^*_\alpha z^*_\alpha )
+ i \lambda_{3} ( P_\alpha z_\alpha - P^*_\alpha z^*_\alpha )\Bigr ] \;\Bigr ),
\end{array} \eqno(3.16)
$$
and subsequently integrate out the momenta as well as the multiplier 
fields.  In the above action, the term proportional to $T_{1}$, occurring 
in the FCC of eqn. (3.14) is absorbed in the $\lambda_{1}$- term.
Since these variables appear linearly, the classical equations of motion 
can be used and one recovers the partition function as
$$
\begin{array}{lcl}
{\cal Z} = {\displaystyle \int}\; {\cal D} (z, z^*)\;
\delta (|Z|^2 - 1)\; \delta (\chi) \;\mbox {det} \{ T, \chi \}\;
\mbox {exp} \Bigl ( i S \Bigr ),
\end{array} \eqno(3.17)
$$
where the classical action $I_{cl}$ is given by the equation (3.1).

In the next Section, we shall take up the Batalin-Tyutin extension
of the above model.\\

\noindent
{\bf 4  Batalin-Tyutin extension }\\

\noindent
The nonlinear nature of the SCC (as well as the presence of the fields
in the SCC constraint algebra) prompts us to rely on the
Batalin-Fradkin-Vilkovisky \cite{bfv} formalism and the Batalin-Tyutin
\cite{bt} scheme. In the latter scheme, the phase space is extended by
incorporating auxiliary fields (also known as B-T fields). In
certain cases, like in the BT extension of the Proca model,
these fields can be identified with the
Stueckelberg field of the Stueckelberg formalism. In fact, the
extension renders the SCC's to FCC's with the advantage that the
Dirac brackets are not required. Furthermore, the path integral 
measure becomes simpler (i.e., no Senjanovic measure is needed) and the gauge
freedom is enhanced so that some convenient gauge conditions can be
introduced in the formalism.

Using the previous results \cite{bbg}, we extend the SCC's as follows

$$
\begin{array}{lcl}
T_{1} \;\;\;\rightarrow\;\;\;\tilde T_{1}
&\equiv& |Z|^2 - 1 + 2 \phi_{1}, \nonumber\\
T_{2} \;\;\;\rightarrow\;\;\;\tilde T_{2}
&\equiv& \Pi_\alpha z_\alpha + \Pi^*_\alpha z^*_\alpha + 2 |Z|^2 \phi_{2},
\end{array}\eqno(4.1)
$$
where $\phi_{i}$ are the B-T fields obeying
$$
\begin{array}{lcl}
\{ \phi_{i} (x), \phi_{j} (y) \} = - \frac{1}{2}
\varepsilon_{ij} \delta (x - y).
\end{array} \eqno(4.2)
$$
Thus, modulo an overall factor, $\phi_{1}$ and $\phi_{2}$ fields can
be treated as a canonical pair. When convenient, we shall denote them by
$ \phi_{1} \equiv \phi$ and $ - 2 \phi_{2} \equiv \Pi_{\phi}$ respectively.
This leads to the Abelianization of the SCC's, i.e.,
$ \{ \tilde T_{1}, \tilde T_{2} \} = 0$. In \cite{bbg}, the corresponding
first-class `extended' Hamiltonian was obtained as an infinite series in
higher powers of the auxiliary variables. In the present model, the
application of the BT scheme (see, e.g. \cite{bbg})  will be even more 
complicated due
to the presence of Hopf term. However, a remarkable extension of the 
above scheme was put forward in  \cite{bfv} where it was proved that there 
exists a one-to-one mapping between the physical variables and an `improved' 
set of variables (appearing as a power series in B-T fields) with the 
property that they commute with the extended SCC's. The complete BT extended 
theory, which is now a gauge theory, can now be
obtained by simply replacing the physical variables in ${\cal H}_{cl}$ 
and the original FCC's by their improved counterpart. This procedure was 
further developed in \cite{kor}  where
it was shown that, at least in the $CP^1$ 
and $O(3)$ nonlinear $\sigma$-models, the infinite series in the extended
Hamiltonian ($ \tilde {\cal H}$) can be summed to give a compact 
expression for the same.

Using the Batalin-Tyutin prescription, one can write the improved 
variables, denoted here by corresponding tildes, for the present case 
as \cite{kor, kor1}
$$
\begin{array}{lcl}
\tilde z_{\alpha} (x) = z_{\alpha} (x) \Bigl [ 1 -
{\displaystyle \sum}_{n = 1}^{\infty} C_{n}^{(z)}
\bigl (\frac{\phi_{1}} {|Z|^2} \bigr )^n
\Bigr ] \equiv z_{\alpha} (x) \;{\cal A},
\end{array} \eqno(4.3)
$$
$$
\begin{array}{lcl}
\tilde \Pi_{\alpha} (x) =
\Bigl [ \Pi_{\alpha} (x) + z_{\alpha}^* (x) \phi_{2} (x)
\Bigr ] \;\Bigl [ 1 + {\displaystyle \sum}_{n}^{\infty} C_{n}^{(\pi)}
\bigl (\frac{\phi_{1}} {|Z|^2} \bigr )^n
\Bigr ] \equiv (\Pi_{\alpha}  + z_{\alpha}^* \phi_{2}) \;{\cal B},
\end{array} \eqno(4.4)
$$
where expressions for $C's$ are
$$
\begin{array}{lcl}
C_{n}^{(z)} &=& ( C_{(n)}^{(z)})^*
= \frac{ (-1)^n (2n - 3)!!}{n!}, \nonumber\\
C_{n}^{(\pi)} &=& (C_{n}^{(\pi)})^*
= \frac{ (-1)^n (2n - 1)!!}{n!}, \nonumber\\
( - 1 ) ! ! &=& 1, \qquad  n ! ! = n (n -2) (n - 4)........
\end{array} \eqno(4.5)
$$
The following useful identities \cite{kor}
$$
\begin{array}{lcl}
{\cal A}\; {\cal B} = 1, \qquad\;\;
( {\cal A} )^2 = ( {\cal B} )^{-2} = \frac{ |Z|^2 + 2 \phi_{1}} 
{|Z|^2}, \end{array} \eqno(4.6)
$$
show that
$$
\begin{array}{lcl}
\tilde T_{1} = |\tilde Z|^2 - 1, \qquad
\tilde T_{2} = \tilde \Pi_\alpha \tilde z_\alpha 
+ \tilde \Pi^*_\alpha \tilde z^*_\alpha.
\end{array} \eqno(4.7)
$$
It is elementary to check that the improved variables commute
with the modified constraints $\tilde T_{i}, (i = 1, 2)$.
A crucial property of the improved variables, proved by Batalin and 
Tyutin in \cite{bt}, is that the ``tilde'' of the products is the product
of tildes, i.e., $  \tilde {(AB)} =   (\tilde A) (\tilde B) $ 
for any two variables $A$ and $B$. This key property allows us to write the 
improved (or first-class) Hamiltonian as:
$$
\begin{array}{lcl}
\tilde H = {\displaystyle \int}\; d^2 x\; \tilde {\cal H } = 
\int d^2 x\; \Bigl ( |\tilde P|^2 + |\partial_{i} \tilde Z|^2 -
| \tilde Z^\dagger \partial_{i} \tilde Z|^2 \Bigr ).
\end{array} \eqno(4.8)
$$
This improved Hamiltonian, by construction, commutes with the
constraints $\tilde T_{1}$ and $\tilde T_{2}$. Note here that since the form
of the Hamiltonian (4.8) here is just the same as that of (3.4), this 
Hamiltonian (4.8) too admits solitonic configurations and associated
topological currents. The corresponding expression for the topological current
can be obtained here by just replacing $Z$ variables in (2.12) by their `images'
$\tilde Z$. We can now do the same for the angular momentum $`J'$. The
improved version $\tilde J$ can be now trivially obtained from $J$. As was
observed in \cite{cm} that, the expressions for angular momentum obtained 
either from Noether's prescription ($J^N$) or through the symmetric expression
for energy-momentum tensor ($J^{(s)}$) (cf. (2.26)) turn out to be identical
thereby indicating the absence of any fractional spin imparted by the Hopf
term at the classical level. The improved $\tilde J$ is therefore given by
$$
\begin{array}{lcl}
\tilde J = {\displaystyle \int}\; d^2 x\;  
\varepsilon_{ij} x_{i} \bigl (\tilde \Pi_{\alpha}
\partial_{j} \tilde z_{\alpha} + \tilde \Pi_{\alpha}^{*} \partial_{j}
\tilde z_{\alpha}^{*} \bigr ).
\end{array} \eqno(4.9)
$$

Before introducing the original first-class constraint $T$,
as given by (3.14), it is worthwhile to comment on the algebra of
the improved variables. In fact, it has been proved in \cite{bt} that 
the following identification between the algebra of physical and improved
variables holds
$$
\begin{array}{lcl}
\{ A, B \}_{DB} = G_{AB}\; \;\;\Rightarrow \;\; \{ \tilde A,  \tilde B 
\}_{PB} = \tilde G_{AB},
\end{array} \eqno(4.10)
$$
where $ \{ A, B \}_{DB} $ is the Dirac bracket between the physical 
variables $A$ and $B$ with the SCC's considered as strong relation, whereas the 
bracket $\{ \tilde A, \tilde B \}_{PB}$ stands for the Poisson bracket in the 
extended phase space. With the SCC's: $ T_{1} = |Z|^2 - 1$ and $ T_{2} = 
z_\alpha \Pi_\alpha + z^*_\alpha \Pi^*_\alpha $, 
it is straightforward to compute the following DB's
$$
\begin{array}{lcl}
\{ z_{\alpha}, z_{\beta} \}_{DB} &=& \{ z_{\alpha}, z_{\beta}^* 
\}_{DB} = 0, \nonumber\\
\{ z_{\alpha},  \Pi_{\beta}^* \}_{DB} &= & - \frac{1} {2 |Z|^2}\;
z_{\alpha} z_{\beta} \delta(x -y), \nonumber\\
\{ z_{\alpha},  \Pi_{\beta} \}_{DB} &= &
\bigl (\delta_{\alpha\beta} - \frac{1} {2 |Z|^2}\;
z_{\alpha} z_{\beta}^* \bigr ) \delta(x -y), \nonumber\\
\{ \Pi_{\alpha},  \Pi_{\beta}^* \}_{DB} &= &
\frac{1} {2 |Z|^2}\; \bigl ( \Pi_{\alpha} z_{\beta} -
\Pi_{\beta}^* z_{\alpha}^* \bigr ) \delta(x -y), \nonumber\\
\{ \Pi_{\alpha},  \Pi_{\beta} \}_{DB} &= &
\frac{1} {2 |Z|^2}\; \bigl ( \Pi_{\alpha} z_{\beta}^* -
\Pi_{\beta} z_{\alpha}^* \bigr ) \delta(x -y),
\end{array}\eqno(4.11)
$$
In fact, as has been shown in \cite{cm} that these structures of the DBs
are inherited from the global $SU(2)$ invariant $S^3$ model.
>From (4.11), as mentioned before, the algebra for the improved variables
are obtained just by replacing each variable by its improved `image', 
e.g.,
$$
\begin{array}{lcl}
\{ \tilde \Pi_{\alpha}, \tilde \Pi_{\beta} \} =
\frac{1} {2 |\tilde Z|^2}\; \bigl (\tilde \Pi_{\alpha} \tilde 
z_{\beta}^* - \tilde \Pi_{\beta} \tilde z_{\alpha}^* \bigr ) \delta(x -y).
\end{array}\eqno(4.12)
$$
Now we are ready to introduce the first-class constraint $\tilde T$ 
in the extended phase space as an `image' of the original FCC
(3.14), using the prescription of [3], as mentioned earlier
$$
\begin{array}{lcl}
T \;\;\Rightarrow \;\;\tilde T
= \tilde T_{3} + 4 \theta i \;\varepsilon_{ij}\;
(\partial_{i} \tilde Z^\dagger \partial_{j} \tilde Z) \; \tilde T_{1}.
\end{array}\eqno(4.13)
$$
Obviously, by construction, we have the following algebra
$$
\begin{array}{lcl}
\{ \tilde T_{1}, \tilde T \} = \{ \tilde T_{2}, \tilde T \} = 0,
\end{array} \eqno(4.14)
$$
and, finally, it can be shown that
$$
\begin{array}{lcl}
\{  T,  {\cal H}_{cl} \}_{DB} = 0,\;\;
\Rightarrow \;\; \{ \tilde T, \tilde H \} = 0, \qquad \tilde 0 = 0.
\end{array} \eqno(4.15)
$$
Thus, our gauge theory constitutes of {\it three Abelian} FCC's $\tilde 
T_{1}, \tilde T_{2}$ and $ \tilde T $ with the first-class Hamiltonian $ \tilde 
H$ in the extended phase space. This Hamiltonian turns out to be in 
involution
with the FCC's. As has been mentioned in the introduction, the presence 
of the original first-class constraint $T$ (3.14) and its corresponding 
extension $\tilde T$ (4.13) has been completely ignored in \cite{kor1}.
These FCCs $\tilde T_{1}, \tilde T_{2}$ in (4.1) and $\tilde T_{3}$
whose explicit form is given as
$$
\begin{array}{lcl}
\tilde T_{3} = i \Bigl (
\Pi_\alpha z_\alpha - \Pi_{\alpha}^* z_{\alpha}^* - 2 \theta \varepsilon_{ij}
({\cal A})^4\; \bigl ( |Z|^2 \partial_{i} Z^\dagger \partial_{j} Z
+ Z^\dagger \partial_{i} Z \partial_{j} Z^\dagger Z \bigr ) \Bigr )
\end{array} \eqno(4.16)
$$
In its infinitesimal form, the gauge transformations generated by 
$\tilde T_{a}$ (a = 1, 2, 3) (4.1), on a generic field $\Phi (x)$, is
$$
\begin{array}{lcl}
\delta_{a} \Phi (x) = {\displaystyle \int} d^2 y f_{a} (y)
\{ \Phi (x), \tilde T_{a} (y) \}, \qquad \mbox {(no summation)}
\end{array} \eqno(4.17)
$$
where $f_{a} (y)$ are some arbitrary 
parameters of transformations and can be taken to be
smooth functions. In certain cases, it may be
necessary to restrict them further to functions having compact supports. A
straightforward calculation yields the following results
$$
\begin{array}{lcl}
\delta_{1} Z (x) &=&  0, \qquad \delta_{1} \phi (x) = 0,
\qquad \delta_{1} \Pi_{\phi} = - 2 f_{1} (x), \nonumber\\
\delta_{2} Z (x) &=&  f_{2} (x) Z(x), \qquad \delta_{2} \phi (x) = 
- f_{2} (x) |Z|^2, \qquad \delta_{2} \Pi_{\phi} = 0, \nonumber\\
\delta_{3} Z(x) &=& i f_{3}(x) Z(x), \qquad \delta_{3} \phi (x) = 0.
\end{array} \eqno(4.18)
$$
Here we have intentionally omitted $\delta_{3} \Pi_{\phi}$ as it is a bit
complicated and will not be very useful for our further discussions. At this
stage, one can make certain observations regarding the nature of the gauge
transformations generated by these FCCs. First note that only $\tilde T_{3}$
(just as the original $T_{3}$ (3.12)) generates $U(1)$ gauge transformation
on the $Z$ fields, whereas under $\tilde T_{2}$ the $Z$ field undergoes a scale
transformation but remains invariant under $\tilde T_{1}$. As far as the BT 
fields are concerned, the field $\Pi_{\phi}$ remains invariant under 
$\tilde T_{2}$ but undergoes a shift under $\tilde T_{1}$. And the field
$\phi (x)$ remains invariant under $\tilde T_{1}$ and $\tilde T_{3}$ but
transforms non-trivially under $\tilde T_{2}$. We shall make use of these facts
to restrict the form of the wave functional of the system in the following
section.

Finally, the explicit expression for the Hamiltonian $ \tilde H = \int d^2 x
\tilde {\cal H}$ can be simplified to a great extent and we obtain:
$$
\begin{array}{lcl}
\tilde  H &=& {\displaystyle \int} \; d^2 x
\Bigl [ \bigl ( |\Pi|^2 + \phi_{2} (\Pi z + \Pi^* z^*) + |Z|^2
(\phi_{2})^2 \bigr )\;\;
\frac{1} { {\cal A}^2} - \partial_{i} {\cal A} \partial_{i} (|Z|^2 {\cal A})
\tilde T_{1} \nonumber\\
&-& \theta \bigl ( \Pi^* M^* + \Pi M \bigr )\; {\cal A}^2 + \theta^2
({\cal A}^2)^3 |M|^2 + |\partial_{i} Z|^2 {\cal A}^2
- |Z^\dagger \partial_{i} Z|^2 ({\cal A}^2)^2 \Bigr ],
\end{array} \eqno(4.19)
$$
where the expression for ${\cal A}^2$ is given by eqn. (4.6). In the
above expression for the Hamiltonian, the total space derivative terms
have been neglected. We can now proceed in an exactly similar manner to
rewrite the improved version of the angular momentum to get
$$
\begin{array}{lcl}
\tilde J = J + {\displaystyle \int}\; d^2 x\; \varepsilon_{ij} x_{i}
\Bigl [ \phi_{2} \partial_{j} (|z|^2) + \partial_{j} (ln A) \tilde T_{2}
\Bigr ].
\end{array} \eqno(4.20)
$$

We would like to point out, at this stage,  that
we have not exploited the ad-hoc and somewhat artificial restriction 
like the so-called conformal gauge condition \cite{kor,kor1}, in the above 
derivation.  This Hamiltonian $\tilde H$ 
and angular momentum $\tilde J$ will now be used to calculate the 
expectation value of the energy- and angular momentum operators 
between physical states in the next section
following a method analogous to the one followed in \cite{no}.\\

\noindent
{\bf 5 Quantum correction to energy and angular momentum}\\

\noindent
In this section, we are going to compute the quantum correction to energy-
and angular momentum 
expectation value obtained by sandwiching the Hamiltonian operator between 
state vectors of the physical Hilbert space ${\cal H}_{ph}$. We shall also 
discuss about its possible physical implications.

To begin with, we shall have to elevate all the three FCC's into three hermitian
operators. At the classical level, these constraints (4.1) and (4.16) are
$$
\begin{array}{lcl}
\tilde T_{1} &=& z_{\alpha}^* z_{\alpha} - 1 + 2 \phi_{1} \approx 0,
\nonumber\\ \tilde T_{2} &=& \Pi_\alpha z_\alpha + \Pi_{\alpha}^* z_{\alpha}^*
+ 2 z_{\alpha}^* z_{\alpha}\; \phi_{2} \approx 0, \nonumber\\
\tilde T_{3} &=& i \Bigl (
\Pi_\alpha z_\alpha - \Pi_{\alpha}^* z_{\alpha}^* - 2 \theta \varepsilon_{ij}
({\cal A})^4\; \Bigl ( |z|^2 \partial_{i} z_{\alpha}^* \partial_{j} z_{\alpha}
+ z_{\alpha}^* \partial_{i} z_{\alpha} \partial_{j} z_{\beta}^* z_{\beta}
\Bigr ) \Bigr ) \approx 0.
\end{array} \eqno(5.1)
$$
Note that here we have written these forms of the constraints in component form
rather than in a matrix form. This is because, at the quantum level, the 
complex conjugation of fields will be replaced by hermitian conjugates of
field operators. And this may create confusion with the hermitian conjugates
$ Z^\dagger = ( z_{1}^*, \; z_{2}^* )$ of the ordinary doublet of two
component field $ Z = \left ( \begin{array}{c} z_{1}\\ z_{2}\\
\end{array} \right )$. These two operations of taking hermitian conjugates
clearly correspond to two distinct spaces. The former is in the Hilbert
space of states (i.e. Fock space) whereas the latter is in the space of fields
$z_{1}$ and $z_{2}$. With this, the hermitian form of the constraints (5.1)
look as
$$
\begin{array}{lcl}
\hat {\tilde T_{1}} &=& \hat z_{\alpha}^\dagger 
\hat z_{\alpha} - 1 + 2 \hat \phi_{1} \approx 0, \nonumber\\
\hat {\tilde T_{2}} &=& \hat \Pi_\alpha \hat z_\alpha 
+ \hat z_{\alpha}^\dagger \hat \Pi_{\alpha}^\dagger 
+ 2 \hat z_{\alpha}^\dagger \hat z_{\alpha}\; 
\hat \phi_{2} \approx 0, \nonumber\\
\hat {\tilde T_{3}} &=& i \Bigl ( \hat \Pi_\alpha \hat z_\alpha 
- \hat z_{\alpha}^\dagger \hat \Pi_{\alpha}^\dagger  
- 2 \theta \varepsilon_{ij} (\hat {\cal A})^4\; 
\Bigl (\hat z_{\beta}^\dagger \hat z_{\beta}
 \partial_{i} \hat z_{\alpha}^\dagger \partial_{j} \hat z_{\alpha}
+ \hat z_{\alpha}^\dagger
 \partial_{i} \hat z_{\alpha} \partial_{j} \hat z_{\beta}^\dagger 
\hat z_{\beta} \Bigr ) \Bigr ) \approx 0.
\end{array} \eqno(5.2)
$$
The basic equal time commutation relations among the fields and their
corresponding conjugate momenta variables are trivially obtained from
their basic Poisson brackets to get
$$
\begin{array}{lcl}
&&[ \hat z_{\alpha} (x), \hat \Pi_{\beta} (y) ] = i 
\hbar \delta_{\alpha \beta}\;\delta (x -y), \nonumber\\
&&[ \hat z_{\alpha}^\dagger (x), \hat \Pi_{\beta}^\dagger (y) ] 
= i \hbar \delta_{\alpha\beta} \delta (x -y), \nonumber\\
&&[ \hat \phi_{\alpha} (x), \hat \phi_{\beta} (y) ] = - \frac {i}{2} 
\hbar \varepsilon_{\alpha \beta}\;\delta (x - y),
\end{array} \eqno(5.3)
$$
and the rest of the brackets vanish. Note that $\hat \phi_{\alpha}$ are
the hermitian operators now. The physical Hilbert space ${\cal H}_{ph}$ is now
given as the kernel of the three FCC's
$$
\begin{array}{lcl}
\hat {\tilde T_{a}} |\Psi>_{ph} = 0, \qquad \forall \;\;a = 1, 2, 3,
\end{array} \eqno(5.4)
$$
which means that the physical states $|\Psi>_{ph}$ are gauge invariant since
$\tilde T_{a}$'s are the generators of the gauge transformations (4.18). This
can restrict the form of the ``physical'' wave functional 
$\Psi [ z_\alpha (x), z^*_\alpha (x), \Pi_{\phi} (x) ] = 
< z_{\alpha} (x), z^*_\alpha (x), \Pi_{\phi} (x)| \Psi >_{ph}$
considerably. Note that we have included here $\Pi_{\phi}$ as one of the 
arguments in the wave functional $\Psi$. Actually, this is a matter of choice
as we could have easily chosen $\phi_{1}$ as conjugate momentum corresponding
to the coordinate variable $\phi_{2}$. They are just related by a trivial
canonical transformation. Also since the BT fields in the extended space 
commute with that of the original $CP^1$ phase space variables ($ z_{\alpha},
\Pi_{\alpha}$), and also due to their transformation properties in (4.18),
the physical Hilbert space ${\cal H}_{ph}$ (which is a subspace of
the total Hilbert space),  can be thought of as a direct product
$ {\cal H}_{1} (z) \times {\cal H}_{2} (\phi)$ or $ {\cal H}_{1} (z)
\times {\cal H}_{2} (\Pi_{\phi})$ and consequently one can easily either
consider ``position'' representation or ``momentum'' representation for
the second Hilbert space ${\cal H}_{2}$ irrespective of the representation
one considers for ${\cal H}_{1}$. Our particular choice here was dictated
by the observation that only $z(x)$ field transforms non-trivially under
$\tilde T_{2}$ but $\Pi_{\phi}$ remains invariant under $\tilde T_{2}$ (4.17).
It is just just other way around for $\tilde T_{1}$ and thus, our analysis
will become simpler.

If we now make a simple demand that the ``physical'' wave functional
$ \Psi (z_{\alpha}, z^*_\alpha, \Pi_{\phi})$ remains invariant under the action
of $\tilde T_{1}$, i.e.,
$$
\begin{array}{lcl}
\Psi (z_{\alpha}, z^*_\alpha, \Pi_{\phi}) =  
\Psi (z_{\alpha}, z^*_\alpha, \Pi_{\phi} + \delta_{1} \Pi_{\phi} ),   
\end{array} \eqno(5.5)
$$
then it follows, using (4.18), that
$$
\begin{array}{lcl}
{\displaystyle \frac{\delta \Psi (z_{\alpha}, z^*_\alpha, \Pi_{\phi})}
{\delta \Pi_{\phi}}} = 0. 
\end{array} \eqno(5.6)
$$
Going to the ``momentum'' representation, where the operator 
$\hat \phi (x) $ is given by 
$  i \hbar \frac{\delta} {\delta \Pi_{\phi}(x)}$, it clearly follows that
$$
\begin{array}{lcl}
\hat {\phi} |\Psi>_{ph} = 0. 
\end{array} \eqno(5.7)
$$
It immediately follows that $|\psi>_{ph}$ must belong to ${\cal H}_{1} \times
\{0\}$, with the second factor consisting of ``zero'' element of ${\cal H}_{2}$
(i.e., $ 0 \in {\cal H}_{2}$) as otherwise one can easily show that the
hermitian operator $ (\hat \phi (x) \hat \Pi_{\phi}(y) 
+ \hat \Pi_{\phi}(y) \hat \phi (x))$ acting on 
$|\Psi>_{ph}$ will produce imaginary eigenvalue ($ i \hbar \delta (x - y)$)
\footnote{ We are assuming that these wave functionals are ``normalizable''
in the functional sense.}. Consequently
$$
\begin{array}{lcl}
\hat \Pi_{\phi} |\Psi>_{ph} = 0. 
\end{array} \eqno(5.8)
$$
We can thus identify ${\cal H}_{ph}$ to be basically isomorphic to 
${\cal H}_{1} (z)$ itself. With this, the above wave functional 
reduces to $ \Psi [ z(x), z^* (x) ]$, 
thus, depending entirely on $z(x)$ and $z^* (x)$.
It is not unexpected, as the condition (5.7) and (5.8) correspond to
the unitary gauge conditions (in the ``weak'' form) to be used in
the next Section. But here we would like to make the following 
observations. Condition (5.4) does not necessarily imply, in general,
that (5.5) has to be satisfied. It can undergo, for example, an
overall scaling transformations so that $|\Psi>_{ph}$ can correspond
to the same element in the projective Hilbert space. 
In fact, one can check that under $\hat {\tilde T_{2}}$, the wave
functional $\Psi$ undergoes a scaling transformation. In this case,
the conditions like unitary gauge (cf. (5.7, 5.8)) (in the ``weak'' 
form) may not hold. But whatever conditions are imposed to define
${\cal H}_{ph}$, it must be isomorphic to what we have found. However,
the unitary gauge will make our computations much simpler in this
Section. 

We are now in a position to compute the ``quantum shift'' in the energy
eigenvalues and study its possible physical significance. For that, we first
need to write the Hamiltonian (4.18) in a hermitian form. But before that,
it will be advantageous to rewrite the classical expression (4.18) itself
in a form suitable for this purpose. Using (4.1) and (4.6), the Hamiltonian
can be written as
$$
\begin{array}{lcl}
\tilde H &=& {\displaystyle \int} \; d^2 x\; \Bigl [
\Bigl ( \frac{ |Z|^2 } { \tilde T_{1} + 1} \Bigr )\;
\Bigl [ |\Pi|^2 - \frac{1}{2} \tilde T_{2} \Pi_{\phi} - \frac{1}{4}
|Z|^2 (\Pi_{\phi})^2 \Bigr ] - \theta \frac{( \tilde T_{1} + 1)} {|Z|^2}
\bigl ( \Pi^* M^* + \Pi M \bigr ) \nonumber\\
&+& \theta^2 \Bigl ( \frac { \tilde T_{1} + 1} { |Z|^2 } \Bigr )^3 |M|^2
+ \Bigl ( \frac { \tilde T_{1} + 1} { |Z|^2 } \Bigr ) |\partial_{i} Z|^2
- \Bigl ( \frac { \tilde T_{1} + 1} { |Z|^2 } \Bigr )^4 
|Z^\dagger \partial_{i} Z|^2 - \partial_{i} {\cal A} \partial_{i}
\bigl ( {\cal A} |Z|^2 \bigr ) \tilde T_{1}
\Bigr ]. 
\end{array} \eqno(5.9)
$$

There is no unique expression for the corresponding hermitian quantum
Hamiltonian as there is a natural ambiguity arising from different and
inequivalent operator orderings. And, inequivalent but consistent operator
orderings give rise to inequivalent quantization. The point we want to
emphasize, at this stage, is the fact that the operator ordering problems
have been gotten rid of by this BT scheme, as the symplectic structure is
now given by the basic commutators obtained by elevating the basic PB 
structure and not the complicated DBs. The only point we shall not worry
about is the divergent nature of the product of the field operators at
the same space-time point.
This is because, at the present level of rigour,
we are only interested in establishing the presence and qualitative nature
of quantum corrections, without going into the quantitative estimates.
For this, we shall try to work with
one of the {\it simplest} and yet nontrivial operator ordering. 

Now coming to the energy expectation value of the Hamiltonian 
operator, we follow the definition 
$$
\begin{array}{lcl}
E = {\displaystyle \frac{_{ph}< \Psi | \hat {\tilde H} | \Psi >_{ph}}
{_{ph}< \Psi | \Psi >_{ph}}}. 
\end{array} \eqno(5.10)
$$
As physical states $ |\Psi>_{ph} $'s are annihilated by the FCC's 
$\hat {\tilde T_{a}}$ (5.4) and also by $\hat \phi (x) $ and $\hat
\Pi_{\phi}$ (5.7, 5.8), we order the various factors in each term
in such a manner that either $\hat {\tilde T_{a}}, \hat \phi $  or
$ \hat \Pi_{\phi}$ appear at the right most or left most place in
each term in $\hat {\tilde H}$. also the hermitian forms of the constraints
(5.2) are kept intact, i.e., the permutations of the field operators
appearing within and without $\hat {\tilde T_{a}}$ are not considered.
Furthermore, the expressions in the denominators, involving field operators,
are re-written in terms of FCC (5.2), so that the expressions involving
the quotients of field operators can be avoided. 
Let us do it term by term. For the 
first term in the integrand of (5.9), we write
$$
\begin{array}{lcl}
\hat {\tilde H_{1}} = \frac{1}{2}\;
{\displaystyle \int} d^2 x\; \Bigl [ \frac{1}{\hat {\tilde T_{1}} + 1}
\Bigl ( \bigl ( \hat z_{\alpha}^\dagger \hat z_{\alpha} \hat 
\Pi_{\beta}^\dagger \hat \Pi_{\beta} \bigr )_{w} - \frac{1}{2}
\bigl ( (\hat z_{\alpha}^\dagger \hat z_{\alpha}
\hat {\tilde T_{2}} \bigr )_{w} \hat \Pi_{\phi} \bigr )
- \frac{1}{4}\; \hat z_{\alpha}^\dagger \hat z_{\alpha} 
\hat z_{\beta}^\dagger \hat z_{\beta} (\hat \Pi_{\phi})^2 \Bigr ) 
+ \mbox {h. c} \Bigl ]
\end{array} \eqno(5.11)
$$
Here the subscript $w$ stands for the Weyl ordering \cite {lee} 
and the composite operators are hermitian
by construction. Further, the Weyl ordering involving the FCC ($\hat {\tilde
T_{2}}$) is performed by treating it as a single object 
as we have mentioned earlier. We could have gone
for simpler ordering than Weyl one for the {\it first} term within the
parenthesis in (5.11) as
$$
\begin{array}{lcl}
\frac{1}{2}\; 
\Bigl (\hat z_{\alpha}^\dagger \hat z_{\alpha} \hat 
\Pi_{\beta}^\dagger \hat \Pi_{\beta} +
\hat \Pi_{\beta}^\dagger \hat \Pi_{\beta} 
\hat z_{\alpha}^\dagger \hat z_{\alpha}  \Bigr ).
\end{array} \eqno(5.12)
$$
But as one can easily see,  this ordering turns out to be rather trivial
in nature. Regarding the second and third terms within the parenthesis 
of (5.11), they clearly vanish by using (5.8) when sandwiched between the
physical states in the numerator of (5.10). Besides, the denominator
$(\hat {\tilde T_{1}} + 1)$ also reduces just to unity. With all this,
the contribution of the term (5.11) in the expectation value simplifies to
$$
\begin{array}{lcl}
E_{1} = {\displaystyle \frac{_{ph}< \Psi| \hat {\tilde H_{1}} |\Psi>_{ph}}
{_{ph}< \Psi| \Psi >_{ph}}}
\equiv {\displaystyle \frac{_{ph}< \Psi |\int d^2 x \bigl (
\hat z_{\alpha}^\dagger \hat z_{\alpha} \hat \Pi_{\beta}^\dagger
\hat \Pi_{\beta} \bigr )_{w} | \Psi >_{ph}}
{_{ph}< \Psi | \Psi >_{ph}}}. 
\end{array} \eqno(5.13)
$$
Now coming to the term linear in the Hopf parameter $\theta$ in (5.9), we
note that it involves a factor $\frac{\tilde T_{1} + 1} {|Z|^2}$. This can
be rewritten as $\frac{\tilde T_{1} + 1} {\tilde T_{1} - 2 \phi +1}$. As is
obvious from the appearances of both numerator and denominator, 
their mutual ordering is really irrelevant
as they commute with each-other. Also since the other function involves
$\Pi_{\alpha}$ and $\Pi_{\alpha}^*$, we write the following ordering for the
$\theta$ dependent term:
$$
\begin{array}{lcl}
\hat {\tilde H_{\theta}} = - \frac{\theta}{4}
{\displaystyle \int} d^2 x\; \Bigl [
\frac{( \hat {\tilde T_{1}} + 1)}{( \hat {\tilde T_{1}} - 2 \hat \phi + 1)}
\bigl ( \hat \Pi_{\alpha}^\dagger \hat M_{\alpha}^\dagger + \hat M_{\alpha}
\hat \Pi_{\alpha} + \hat M_{\alpha}^\dagger 
\hat \Pi_{\alpha}^\dagger + \hat \Pi_{\alpha} \hat M_{\alpha} \bigr ) 
+  \mbox {h. c.} \Bigr ].
\end{array} \eqno(5.14)
$$
Again using (5.4) and (5.7), it is clear that when $\hat {\tilde H_{\theta}}$ is 
sandwiched between two physical states, this above mentioned factor,
 reduces to unity effectively. The 
corresponding contribution to the expectation value thus becomes:
$$
\begin{array}{lcl}
E_{\theta} =
{\displaystyle \frac{_{ph}< \Psi | - \frac{\theta}{2}  \int d^2 x 
\bigl ( \hat \Pi_{\alpha}^\dagger \hat M_{\alpha}^\dagger + 
\hat M_{\alpha} \hat \Pi_{\alpha} + \hat M_{\alpha}^\dagger \hat 
\Pi_{\alpha}^\dagger + \hat \Pi_{\alpha} \hat M_{\alpha}
\bigr ) | \Psi >_{ph}}
{_{ph}< \Psi | \Psi >_{ph}}}. 
\end{array} \eqno(5.15)
$$
In rest of the terms in (5.9), there is no need for operator ordering
as all the variables commute among themselves. Consequently, their contribution
to the energy is given by
$$
\begin{array}{lcl}
E_{rest} = {\displaystyle \frac{_{ph}< \Psi | \int d^2 x \bigl (
\theta^2 |\hat M|^2 + |\partial_{i} \hat Z|^2 - |\hat Z^\dagger 
\partial_{i} \hat Z|^2 \bigr ) | \Psi >_{ph}} {_{ph}< \Psi | \Psi >_{ph}}}. 
\end{array} \eqno(5.16)
$$
Thus, the total energy eigenvalue is obtained by adding (5.11), (5.13)
and (5.14):
$$
\begin{array}{lcl}
E &=& E_{1} + E_{\theta} + E_{rest}\nonumber\\
&\equiv & \frac{_{ph}< \Psi |\int d^2 x \Bigl [ \bigl (
\hat z_{\alpha}^\dagger \hat z_{\alpha} \hat \Pi_{\beta}^\dagger
\hat \Pi_{\beta} \bigr )_{w} - \frac{\theta}{2} \bigl (
\hat \Pi_{\alpha}^\dagger \hat M_{\alpha}^\dagger + 
\hat M_{\alpha} \hat \Pi_{\alpha}  +
\hat M_{\alpha}^\dagger \hat \Pi_{\alpha}^\dagger + 
\hat \Pi_{\alpha} \hat M_{\alpha} \bigr ) +
\theta^2 |\hat M|^2 + |\partial_{i} \hat Z|^2 
- |\hat Z^\dagger \partial_{i} \hat Z|^2 \Bigr ]
| \Psi >_{ph}} {_{ph}< \Psi | \Psi >_{ph}}. 
\end{array} \eqno(5.17)
$$
Clearly, we will have to carry out the Weyl ordering only in the first term
in the integrand as indicated in the above equation. The second $\theta$
dependent term is already ordered properly as $\hat M_{\alpha}$ involves
only $\hat z_{\alpha}$ and $\hat z_{\alpha}^\dagger$ variables. Let us now
Weyl order the first term $ \int d^2 x\; \bigl [ \hat z_{\alpha}^\dagger
\hat z_{\alpha} \hat \Pi_{\beta}^\dagger \hat \Pi_{\beta} \bigr ]$. This
can be rewritten as
$$
\begin{array}{lcl}
\int d^2 x \;d^2 y \bigl ( \hat z_{\alpha}^\dagger (x)
\hat z_{\alpha}(x) \hat \Pi_{\beta}^\dagger (y) \hat 
\Pi_{\beta} (y) \bigr )_{w} \delta (x - y).
\end{array} \eqno(5.18)
$$

For the sake of convenience, let us denote, for the time being, the 
integrand in (5.18) (excluding $\delta (x -y)$) just as
$$
\begin{array}{lcl}
\Bigl (\hat z_{\alpha}^\dagger \hat z_{\alpha} \hat \Pi_{\beta}^\dagger
\hat \Pi_{\beta} \Big )_{w},
\end{array} \eqno(5.19)
$$
where $\alpha$ and $\beta$ indices are now taken to include space
indices $x$ and $y$ respectively. In this compact notation, the first 
two non-vanishing commutators in (5.3) can be expressed as
$$
\begin{array}{lcl}
[ \hat z_{\alpha}, \hat \Pi_{\beta} ]
= [ \hat z_{\alpha}^\dagger, \hat \Pi_{\beta}^\dagger ] 
= i \hbar \delta_{\alpha \beta}.
\end{array} \eqno(5.20)
$$
Considering all possible permutations of the variables in (5.19), this
can be Weyl ordered as
$$
\begin{array}{lcl}
\Bigl (\hat z_{\alpha}^\dagger \hat z_{\alpha} \hat \Pi_{\beta}^\dagger
\hat \Pi_{\beta} \Big )_{w} &=& \frac{1}{24}\; \Bigl [ \;4\; \{
\hat z_{\alpha}^\dagger \hat z_{\alpha},  \hat \Pi_{\beta}^\dagger
\hat \Pi_{\beta} \}_{s} 
+  4 \{ \hat z_{\alpha}^\dagger \hat \Pi_{\beta}, 
\hat z_{\alpha} \hat \Pi_{\beta}^\dagger \}_{s} \nonumber\\
&+& \{ \hat z_{\alpha}^\dagger, \hat \Pi_{\beta}^\dagger \}_{s}
\{ \hat z_{\alpha}, \hat \Pi_{\beta} \}_{s} +
\{ \hat z_{\alpha}, \hat \Pi_{\beta} \}_{s} \{ \hat z_{\alpha}^\dagger,
\hat \Pi_{\beta}^\dagger \}_{s} \;\Bigr ],
\end{array} \eqno(5.21)
$$
where the symmetric bracket  between two operators $\hat A$ and $\hat B$
is defined as $ \{ \hat A, \hat B \}_{s} = \hat A \hat B + \hat B \hat A$.
Repeated application of (5.20) allows one to simplify (5.21) as
$$
\begin{array}{lcl}
\Bigl (\hat z_{\alpha}^\dagger \hat z_{\alpha} \hat \Pi_{\beta}^\dagger
\hat \Pi_{\beta} \Big )_{w}
= \hat z_{\alpha}^\dagger \hat z_{\alpha} \hat \Pi_{\beta}^\dagger
\hat \Pi_{\beta} + \frac{\hbar^2}{4}\; \delta_{\alpha\beta}
\delta_{\alpha\beta} - \frac{i \hbar}{2} \delta_{\alpha\beta}
\bigl (\hat z_{\alpha}^\dagger \hat \Pi_{\beta}^\dagger 
+ \hat \Pi_{\beta} \hat z_{\alpha}
\bigr ).  \end{array} \eqno(5.22)
$$
Restoring the continuous spacetime indices, (5.20) can be expressed as
$$
\begin{array}{lcl}
\Bigl (\hat z_{\alpha}^\dagger (x) \hat z_{\alpha} (x)
\hat \Pi_{\beta}^\dagger (y) \hat \Pi_{\beta} (y) \Big )_{w}
&=& \hat z_{\alpha}^\dagger (x) \hat z_{\alpha} (x)
\hat \Pi_{\beta}^\dagger (y) \hat \Pi_{\beta} (y) 
+ \frac{\hbar^2}{4}\; \delta_{\alpha\beta} 
\delta_{\alpha\beta} \delta (x - y) \delta (x -y) \nonumber\\
&-& \frac{i \hbar}{2} \delta_{\alpha\beta} \delta (x - y)
\bigl (\hat z_{\alpha}^\dagger (x) \hat \Pi_{\beta}^\dagger (y) 
+ \hat \Pi_{\beta} (y) \hat z_{\alpha} (x) \bigr ).  \end{array} 
$$
which on further simplification yields
$$
\begin{array}{lcl}
\Bigl (\hat z_{\alpha}^\dagger (x) \hat z_{\alpha} (x)
\hat \Pi_{\beta}^\dagger (y) \hat \Pi_{\beta} (y) \Big )_{w}
&=& \bigl (\hat {\tilde T_{1}} (x) +1 - 2 \hat \phi (x) \bigr )
\hat \Pi_{\beta}^\dagger (y) \hat \Pi_{\beta} (y) 
+ \frac{\hbar^2}{2}\; \bigl (\delta (x -y) \bigr )^2 \nonumber\\
&-& \frac{i \hbar}{2} 
\bigl (\hat {\tilde T_{2}} (x) 
+ \hat z_{\alpha}^\dagger (x) \hat z_{\alpha} (x)
\hat \Pi_{\phi} (x) \bigr ).  \end{array} \eqno(5.23)
$$
So when this term is sandwiched between physical states $|\Psi>_{ph}$ in 
(5.17), the term linear in $\hbar$ effectively drops out and the first term
effectively reduces to $ \hat \Pi_{\beta}^\dagger \hat \Pi_\beta$ as can be
easily seen from (5.4), (5.7) and (5.8). So far, this term $\bigl (
\hat z_{\alpha}^\dagger \hat z_\alpha \hat \Pi_{\beta}^\dagger \Pi_{\beta}
\bigr )_{w}$ yields an ${\cal O} (\hbar^2)$ quantum correction. Now coming to
the term linear in $\theta$ in (5.17), we note that the integral
$\int d^2 x\; \bigl ( \hat \Pi_\alpha (x) \hat M_\alpha (x) \bigr )$ appearing
there can be re-expressed as
$$
\begin{array}{lcl}
{\displaystyle \int} d^2 x d^2 y \hat \Pi_\alpha (y) \hat M_\alpha (x)
\delta (x - y), 
\end{array} \eqno(5.24)
$$
as was done for the case of the first term in (5.18). Again repeated 
application of the basic commutation relation (5.3) allows one to rewrite
($ \int d^2 x \hat \Pi_\alpha \hat M_\alpha (x)$) as
$$
\begin{array}{lcl}
{\displaystyle \int} d^2 x  \hat \Pi_\alpha (x) \hat M_\alpha (x) &=&
{\displaystyle \int} d^2 x \hat M_{\alpha} (x) \hat \Pi_{\alpha} (x)
- i \hbar {\displaystyle \int} d^2 x d^2 y \delta (x - y) \varepsilon^{ij}
\partial^{(x)}_{i} \bigl ( \hat z_{\alpha}^\dagger \hat z_{\alpha} \bigr )
\partial^{(x)}_{j} \delta (x - y) \nonumber\\
&+& i \hbar {\displaystyle \int} d^2 x d^2 y \bigl ( \delta (x - y) \bigr )^2
\varepsilon^{ij} \partial^{(x)}_{i} \hat z_{\alpha}^\dagger \partial^{(x)}_{j}
\hat z_{\alpha}. 
\end{array} \eqno(5.25)
$$
We, therefore, have
$$
\begin{array}{lcl}
{\displaystyle \int} d^2 x  \bigl (\hat \Pi_\alpha (x) \hat M_\alpha (x) +
\hat M_{\alpha}^\dagger (x) \hat \Pi_{\alpha}^\dagger (x) \bigr ) &=&
{\displaystyle \int} d^2 x \bigl (\hat M_{\alpha} (x) \hat \Pi_{\alpha} (x)
+ \hat \Pi_{\alpha}^\dagger (x) \hat M_{\alpha}^\dagger (x) \bigr )
\nonumber\\
&+& 2 i \hbar {\displaystyle \int} d^2 x d^2 y \bigl ( \delta (x - y) \bigr )^2
\varepsilon^{ij} \partial^{(x)}_{i} \hat z_{\alpha}^\dagger \partial^{(x)}_{j}
\hat z_{\alpha}. 
\end{array} \eqno(5.26)
$$
Note that here the second term in (5.25) involving the derivative of the delta
function drops out, as it is skew-hermitian. Now we can write the term 
involving `$\hbar$' in (5.26) in more compact form, using 
$\hat {\tilde T_{1}} |\Psi>_{ph} = 0$ (5.4) and (5.7) to note that $ (\hat 
z_{\alpha}^\dagger \hat z_{\alpha} - 1) |\Psi>_{ph} = 0$. Thus, 
as far as the actions
of the second `$\hbar$' dependent term in (5.24) on $|\Psi>_{ph}$ states,
taken to be an eigen state $| z_\alpha (x)>$ of the
field operator $\hat z_{\alpha}$ as: $\hat z_{\alpha} (x) |z_{\alpha} (x)> 
= z_\alpha (x) | z_\alpha (x)>$, are concerned
\footnote{ A typical eigen state $|z_\alpha (x)>$ can be thought of as 
given by the configuration (2.18) in the $ Q = 1$ sector.}, the integrand 
(up to a factor) can be identified with the topological density $ j^{0}$
(2.12), and allows one, using (2.10), to rewrite (2.26) as
$$
\begin{array}{lcl}
{\displaystyle \int} d^2 x  \bigl (\hat \Pi_\alpha (x) \hat M_\alpha (x) +
\hat M_{\alpha}^\dagger (x) \hat \Pi_{\alpha}^\dagger (x) \bigr ) &=&
{\displaystyle \int} d^2 x \bigl (\hat M_{\alpha} (x) \hat \Pi_{\alpha} (x)
+ \hat \Pi_{\alpha}^\dagger (x) \hat M_{\alpha}^\dagger (x) \bigr ) \nonumber\\
&-& 4\pi \hbar {\displaystyle \int} d^2 x d^2 y \bigl ( \delta (x - y) \bigr )^2
\hat j^{0} (x), \nonumber\\
&=& {\displaystyle \int} d^2 x \bigl (\hat M_{\alpha} (x) \hat \Pi_{\alpha} (x)
+ \hat \Pi_{\alpha}^\dagger (x) \hat M_{\alpha}^\dagger (x) \bigr ) \nonumber\\
&-& 4\pi \hbar Q {\displaystyle \int}  d^2 y \bigl ( \delta (x - y) \bigr )^2
\end{array} \eqno(5.27)
$$
Thus, ultimately, using (5.17),
the energy expectation value of $E$ boils down to 
$$
\begin{array}{lcl}
E &=& {\displaystyle \frac{1} {_{ph}< \Psi | \Psi >_{ph}}} \;
_{ph}< \Psi |\int d^2 x \bigl [ \hat \Pi_{\alpha}^\dagger \hat \Pi_{\alpha} 
- \theta (\hat \Pi_{\alpha}^\dagger \hat M_{\alpha}^\dagger 
+ \hat M_{\alpha} \hat \Pi_{\alpha} ) \nonumber\\ &+& \theta^2 
\hat M_{\alpha}^\dagger \hat M_{\alpha} + |\partial_{i} \hat Z|^2 
- |\hat Z^\dagger \partial_{i} \hat Z|^2 \bigr ] | \Psi >_{ph}
\nonumber\\ &+& \frac{\hbar^2}{2}\; {\displaystyle \int}\; d^2 x d^2 y
\bigl ( \delta (x - y) \bigr )^3 + 2 \pi \hbar \theta Q {\displaystyle \int}
d^2 y\; \bigl ( \delta (x -y) \bigr )^2. 
\end{array} \eqno(5.28)
$$

At this stage, one can use $\Pi_{\alpha} = (P_{\alpha} 
+ \theta M_{\alpha}^\dagger)$ and $ \Pi_{\alpha}^\dagger 
= (P_{\alpha}^\dagger + \theta M_{\alpha})$, the quantum
version of (3.4) to show that the first three terms in the integral of (5.28)
simplifies considerably to yield $ P_{\alpha}^\dagger P_{\alpha} =
\Pi_{\alpha}^\dagger \Pi_{\alpha} - \theta (\Pi_{\alpha}^\dagger 
M_{\alpha}^\dagger + M_{\alpha} \Pi_{\alpha}) + \theta^2 
M_{\alpha}^\dagger M_{\alpha}$, so that the entire first term in (5.28)
can be written as:
$$
\begin{array}{lcl}
\bar E = {\displaystyle \frac{_{ph}< \Psi |\int d^2  x \bigl [
\hat P_{\alpha}^\dagger \hat P_{\alpha} +
|\partial_{i} \hat Z|^2 - |\hat Z^\dagger \partial_{i} \hat Z|^2 \bigr ] 
| \Psi >_{ph}}
{_{ph}< \Psi | \Psi >_{ph}}}. 
\end{array} \eqno(5.29)
$$
With this (5.28) can be expressed more compactly as
$$
\begin{array}{lcl}
E = \bar E + \frac{\hbar^2}{2}\; 
{\displaystyle \int}\; d^2x d^2 y \bigl (\delta (x - y) 
\bigr)^3 + 2 \pi \hbar \theta Q \int d^2 y\; \bigl ( \delta (x - y) \bigr )^2.
\end{array} \eqno(5.30)
$$
As one can easily recognize that the integral (5.29) just corresponds to the
Hamiltonian of pure $CP^1$ model. This also happens for vanishing
Hopf term $(\theta = 0)$. We can therefore identify, with some justification, 
$\bar E$ in (5.30) as the classical expression and the second
($ O(\hbar^2) $ term) and the third ($ {\cal O} (\hbar)$ term) in (5.30) 
as quantum corrections. It should be
noted, however, that these terms are highly singular and to extract any 
meaning from these, we have to regularize them. For this purpose, let us
use the Gaussian representation of the two-dimensional delta- function
$$
\begin{array}{lcl}
\delta_{\sigma} (x) = \frac{1} {4 \pi \sigma^2}\; 
e^{- \frac{x^2}{\sigma^2}}.
\end{array} \eqno(5.31)
$$
This represents Dirac-delta function in the limit $ \sigma \rightarrow 0$.
Thus, the first quantum correction can be written as
$$
\begin{array}{lcl}
E^{(1)}_{quan} &=& \frac{\hbar^2}{2}\;{\displaystyle \int}\; d^2x\;d^2y\;
\bigl (\delta (x -y)\bigr )^3, \nonumber\\
&=& \frac{\hbar^2}{2}\; \mbox { Lim}_{\sigma \rightarrow 0}
{\displaystyle \int}\; d^2x\;d^2y\;
\frac{1}{(4\pi\sigma^2)^3}\; e^{-\frac{3 (x -y)^2}{\sigma^2}}.
\end{array} \eqno(5.32)
$$
If we perform the $y$-integration first by translating $y$ appropriately,
then this integral becomes essentially independent of $x$ and upon the
second integration over $x$, the integral diverges. Introducing an area
cut-off $A$ for this $x$-integration, (5.31) can be rewritten,
after some algebra, as
$$
\begin{array}{lcl}
E^{(1)}_{quan} = \mbox {Lim}_{\sigma\to 0 \atop A \to \infty}\;\;
{\displaystyle \frac{\hbar^2\; A}{96\;\pi^2\;\sigma^4}}.
\end{array} \eqno(5.33)
$$
Looking at it dimensionally, it is clear that a quantity having length
dimension should appear in the numerator, which we have taken effectively
to have unit magnitude right in the Lagrangian (2.4). This is clearly highly
divergent and the situation does not improve by considering the corresponding
energy density, i.e., energy per unit area ($ \frac{ E^{(1)}_{quant}}{A}$).
However, the situation is slightly better with the ${\cal O} (\hbar)$ term
in (5.30); namely,
$$
\begin{array}{lcl}
E^{(2)}_{quan} = 2 \pi \hbar \theta Q
{\displaystyle \int} d^2 y\; \bigl ( \delta (x - y) \bigr )^2.
\end{array} \eqno(5.34)
$$
Again using the same regularization (5.31), (5.34) simplifies to
$$
\begin{array}{lcl}
E^{(2)}_{quan} = \mbox {Lim}_{ \sigma \rightarrow 0} \Bigl (
{\displaystyle \frac {1}{4 \sigma^2}} \hbar \theta Q \Bigr ).
\end{array} \eqno(5.35)
$$
Although this is also divergent, the corresponding contribution of the
Hopf term to the energy density
$$
\begin{array}{lcl}
\mbox {Lim}_{ A \rightarrow \infty} 
\frac {E^{(2)}_{quan}} {A} = \mbox {Lim}_{ \sigma \to 0 \atop
A \to \infty} \Bigl (
{\displaystyle \frac {1}{4 A \sigma^2}} \hbar \theta Q \Bigr ),
\end{array} \eqno(5.36)
$$
can be made finite. This is a nontrivial result considering the fact
that the Hopf term is topological in nature. However, note that the quantity
$\mbox{Lim}_{\sigma \rightarrow 0, A \rightarrow \infty}
\bigl (A \sigma^2 \bigr )$, which can
be taken to be finite, is regularization scheme dependent. Nevertheless, 
this result indicates that a generic topological term may contribute 
non-trivially in the energy momentum tensor at the quantum level in the
nontrivial topological sector. 
This point deserves further careful investigation if one is interested
in the explicit numerical estimate of energy.

We now turn our attention to the angular momentum operator. Proceeding exactly 
in the manner, as we have done for energy, we can write down a hermitian form 
of the improved angular momentum (4.20) as
$$
\begin{array}{lcl}
\hat {\tilde J} = \hat J + \frac{1}{2}\; {\displaystyle \int} d^2 x\;
\varepsilon_{ij} \;x_{i}\;\Bigl [ \partial_{j} (ln A)\; \hat {\tilde T_{2}}
+ \hat {\tilde T}_{2} \partial_{j} (ln A) - \hat {\Pi_{\phi}} (x)
\partial_{j} (|\hat z|^2) \Bigr ],
\end{array} \eqno(5.37)
$$
where
$$
\begin{array}{lcl}
\hat J = {\displaystyle \int}\; d^2 x\;
\varepsilon_{ij} \;x_{i}\;\Bigl [
 \hat \Pi_{\alpha} (x)
\partial_{j} \hat z_{\alpha} (x) +  \mbox {h. c.} \Bigr ],
\end{array} \eqno(5.38)
$$
is the original expression of the angular momentum for (3.1), obtained
either by Noether's prescription or through the symmetric energy momentum 
tensor (2.26) and is nothing but the orbital angular momentum. Here also
one starts by changing all the variables to their ``improved'' BT extended [5]
form and algebraic simplification leads to the conventional form (4.20) for
which (5.37) provides the hermitian counterpart.  Clearly,
juxtaposed between physical states $|\Psi>_{ph}$, the extra terms in
(5.37) vanishes as can be easily seen on using (5.4) and (5.8). One
therefore concludes that, unlike the case of energy, there is no quantum
correction in the case of angular momentum. It was argued in \cite{cm}
that Hopf term does not contribute to fractional spin and, now in absence
of any quantum correction, there is no contribution to fractional spin of
quantum mechanical origin either. This is consistent with the collective 
coordinate quantization carried out in Section 2.  \\

\noindent
{\bf 6 BRST quantisation}\\

\noindent
In this Section, we briefly outline the BRST quantisation
for the BT extended  $CP^1$ model coupled to the Hopf term in the
framework of Hamiltonian formalism \cite{bfv}. Since the extension has
already converted the system into a completely first-class system, the
procedure of BRST quantisation is straightforward. One has to introduce
the following three canonical pairs of ghosts, anti-ghosts and
multiplier fields in the Batalin-Fradkin-Vilkovisky scheme

$$
\begin{array}{lcl}
(C^i, \bar P_{i}), \quad (P^i, \bar C_{i}), \quad
(q^i, p_{i}), \quad i= 1, 2, 3,
\end{array}\eqno(6.1)
$$
satisfying the super-Poisson algebra
$$
\begin{array}{lcl}
\{ C^i (x, t), \bar P_{j} (y, t) \} =
\{ P^i (x, t), \bar C_{j} (y, t) \} =
\{ q^i (x, t), p_{j} (y, t) \} = \delta^i_{j}\; \delta (x - y),
\end{array} \eqno(6.2)
$$
where the super-bracket between two variables $A$ and $B$ is defined as
$$
\begin{array}{lcl}
\{ A, B \} =
{\displaystyle {\frac{ \delta A} {\delta q}} |_{r}\; 
{\frac{\delta B} {\delta p}} |_{l} - (-1)^{\eta_{A} \eta_{B}}
{\frac{ \delta A} {\delta p}} |_{l}\; {\frac{\delta B} {\delta q}} |_{r}}.
\end{array} \eqno(6.3)
$$
Here the subscripts $l$ and $r$ stand for the left- and right 
derivatives respectively and $\eta_{A}$ corresponds to the ghost 
number associated with the variable $A$. The Hamiltonian path-integral
for the partition function ${\cal Z}$, is finally written as
$$
\begin{array}{lcl}
{\cal Z} = {\displaystyle \int} {\cal D}  [\;\mu \; ] \;
exp \;\Bigl ( i {\displaystyle \int} d^{3} x
\Bigl [\; \Pi \dot z + \Pi^* \dot z^* + \Pi_{\phi} \dot \phi
+ p_{i} \dot q^i + \bar P_{i} \dot C^i + \bar C_{i} \dot P^i
- {\cal H}_{U}\; \Bigr ] \Bigr ),
\end{array} \eqno(6.4)
$$
where the measure $ {\cal D}  [ \mu ]$ consists of all
the phase (conjugate) variables and ${\cal H}_{U}$ is defined as
the unitarizing Hamiltonian
$$
\begin{array}{lcl}
H_{U} = {\displaystyle \int} \;d^2 x \; {\cal H}_{U} \equiv
{\displaystyle \int} d^2 x \;{\cal H}_{BRST} + \{ \Psi, Q_{B} \}.
\end{array}\eqno(6.5)
$$
In the above equation, the gauge-fixing fermion $\Psi$, the BRST charge
$Q_{B}$ and the BRST Hamiltonian ${\cal H}_{BRST}$ are defined  in a
conventional way \cite{bfv},
$$
\begin{array}{lcl}
\Psi = {\displaystyle \int} d^2 x \bigl ( \bar C_i \chi^i
+ \bar P_{i} q^i \bigr ), \quad
Q_{B} = {\displaystyle \int} d^2 x \bigl ( C^i \tilde T_{i} +
P^i p_{i} \bigr ),
\end{array} \eqno(6.6)
$$
$$
\begin{array}{lcl}
\{ {\cal H}_{BRST}, Q_{B} \} = 0, \quad
\{ Q_{B}, Q_{B} \} = 0.
\end{array} \eqno(6.7)
$$
Here $\chi^i$'s are the three gauge-fixing functions (to be specified 
later) with the restriction that the Poisson-Bracket matrix, consisting of PB's 
among $\tilde T_{i}$ and $\chi^j$, should be invertible. In the construction 
of the BRST invariant Hamiltonian ${\cal H}_{BRST}$, the presence of the
original FCC $T$ (and its improved version $\tilde T$)
creates extra complications. Notice that the improved variables
$ \tilde z, \tilde \Pi, \tilde z^*, \tilde \Pi^*$ were tailored to 
commute with the original SCC's $\tilde T_{1}$ and $\tilde T_{2}$
and hence by construction,
$$
\begin{array}{lcl}
\{ \tilde T_{1}, \tilde {\cal H} \} =
\{ \tilde T_{2}, \tilde {\cal H} \} = 0.
\end{array}
$$
However, this is not in general true for $\tilde T$ and, therefore, in 
general $ \tilde {\cal H} \neq {\cal H}_{BRST} $. Thus,
in an arbitrary model, further modifications
are required  to convert $\tilde {\cal H}$ to ${\cal H}_{BRST}$. But, it
can be explicitly checked that in the present theory
$$
\begin{array}{lcl}
\{ \tilde T, \tilde {\cal H} \} =  0, \quad \Rightarrow \quad
\tilde {\cal H}  = {\cal H}_{BRST},
\end{array}
$$
which leads to
$$
\begin{array}{lcl}
H_{U} = {\displaystyle \int}\; d^2 x \;
\Bigl [ \;\tilde {\cal H} + q^i \tilde T_{i} + \bar P_{i} P^i + p_{i} 
\chi^i + {\displaystyle \int} d^2 y \;
\bar C_{i} (x)\; \{ \chi^i (x), \tilde T_{j} (y) \}\; C^j (y)\; \Bigr ].
\end{array} \eqno(6.8)
$$
In this context, we would like to comment that the introduction of
terms proportional to the FCC's to the improved Hamiltonian
$\tilde {\cal H}$ as in \cite{kor} with the sole purpose of maintaining
the original constraint algebra in the BT scheme is unnecessary and
also seems to be redundant since these constraints already appear in the
action coupled to the {\it arbitrary} multiplier fields. Furthermore,
for the more complicated models such as the one presented here, the form
invariance of the original constraint algebra in the extended phase 
space is not possible.

Our aim here is to construct the Lagrangian in the extended
velocity phase space which implies that all the momenta variables should
be integrated out from (5.4). Fortunately, in our case, the 
momenta appear either linearly or at the most quadratically, and hence the 
classical equations of motion can be used to eliminate them. First of all,
for simplicity, let us note that the
term $ \int d^2x \bigl ( p_{1} \dot q^1 + \bar P_{1} \dot C^1 \bigr )$ 
in the action is dropped because it is a BRST exact piece:
$$
\begin{array}{lcl}
{\displaystyle \int} d^2 x\; \bigl ( p_{1} \dot q^1 + \bar P_{1} \dot 
C^1
\bigr ) = \{ Q_{B}, {\displaystyle \int} d^2 x \bar C_{1} \dot q^1 \}.
\end{array}\eqno(6.9)
$$
This allows us to trivially integrate out the variables $ q^1, p_{1},
P^1, \bar P_{1}, P^2, \bar P_{2}$ leading to the partition function
$$
\begin{array}{lcl}
{\cal Z} &=& {\displaystyle \int}\; {\cal D}  [ \mu  ]
\;\mbox {exp} \;\Bigl ( i {\displaystyle \int} d^3 x\;
\Bigl [ \;\Pi \dot z + \Pi^* \dot z^* + \Pi_{\phi} \dot \phi
+ p_{2} \dot q^2 + p_{3} \dot  q^3 - \bar P_{3} P^3 + \bar P_{3} \dot 
C^3
+ \bar C_{3} \dot P^3 \nonumber\\
&-& q^2 \tilde T_{2} - q^3 \tilde T_{3} - p_{2} \chi^2 - p_{3} \chi^3
- \tilde {\cal H} - {\displaystyle \int} d^2 y\;
\bar C_{i} \{ \chi^i (x), \tilde T_{j} (y) \} C^j (y)
\Bigr ] \Bigr ), \nonumber\\
{\cal D} \bigl [ \mu \bigr ] &=&
{\cal D} \bigl ( \Pi, z, \Pi^*, z^*, \Pi_{\phi}, \phi, q^2, p_{2},
q^3, p_3, C^i, \bar C_i, P^3, \bar P_{3} \bigr )\; \delta (\tilde T_{1})
\;\delta (\chi^1).
\end{array}\eqno(6.10)
$$
In the same way as the above, removal of other BRST exact terms from the
action is not desirable since the other constraints contain the momenta
variables. Presence of $\delta (\tilde T_{1})$ in the measure simplifies
the subsequent calculations considerably since this allows us to write
$$
\begin{array}{lcl}
\tilde T \equiv \tilde T_{3} = (\Pi z) - (\Pi^* z^*)
+ 2 \theta \varepsilon_{ij} ({\cal A}^2)^2
\Bigl [ |Z|^2 \partial_{j} Z^\dagger \partial_{i} Z 
+ (\partial_{i} Z^\dagger \; Z)
(Z^\dagger \partial_{j} Z) \Bigr ].
\end{array}
$$
So far, the choice of gauge has remained arbitrary. Let us choose
a unitary gauge

$$
\begin{array}{lcl}
\chi^{1} = T_{1} (z, z^*), \qquad
\chi^{2} = T_{2} (z, z^*, \Pi, \Pi^*).
\end{array} \eqno(6.11)
$$
This choice, at least, ensures that the extension related to the BRST
procedure reproduces the original action (3.1) in the limit when BT
auxiliary fields vanish. Notice that $\chi^{3}$ is still kept 
arbitrary.
However, for simplicity, let us restrict $\chi^{3}$ to comprise only
of $z$ and $z^*$ fields and $\{ \tilde T_{2}, \chi^{3} \} = 0$. This
finally leads to the following condition
$$
\begin{array}{lcl}
\{ \chi^{3} (x), \tilde T_{3} (y) \} = \sigma (x, y) \neq 0.
\end{array} \eqno(6.12)
$$
In this particular set up, we are allowed to remove the BRST exact
term $\int \; d^2 x\; \bigl ( p_{3} \dot q^3 + \bar C_{3} \dot P^3 \bigr 
) = \{ Q_{B}, \int d^2 x\; \bar C_{3} \dot q^3 \} $  from the action and
similar analysis as done earlier changes the measure to
$$
\begin{array}{lcl}
{\cal D} \; [ \mu  ]\; = {\cal D} \bigl (z, z^*, \Pi , \Pi ^* ,
\phi, \Pi _\phi , p_2, C_{i}, \bar C^i \bigr )\; \delta ( \tilde 
T_{1})\;
\delta ( \chi^1 ) \;\delta (\chi^3).
\end{array} \eqno(6.13)
$$

Up to this point, the effective action is still BRST invariant since we
are still in the BT extended phase space. Now,  the momenta variables
$\Pi$ and $\Pi^*$ can be easily removed by exploiting the equations of
motion. For instance, the following expression
$$
\begin{array}{lcl}
\Pi^*_{\alpha}  = \dot z_{\alpha}
- \bigl ( q^2 + q^3 + p_2 + \frac{1} {2 {\cal A}^2} \Pi_{\phi} \bigr )
z_{\alpha} - \theta\; {\cal A}^2 M^*_{\alpha},
\end{array} \eqno(6.14)
$$
arises from the equations of motion w.r.t. $\Pi_{\alpha}$,
which can be used to eliminate $\Pi^*_\alpha $. This brings
about two Gaussian path-integrals in the remaining momenta variables,
i.e., $p_{2}$ and $ \Pi_{\phi}$. The above integrations produce a long 
expression
for the effective action which is not given here. Let us now simplify 
the
above action further by using the following $\delta$-functions:
$$
\begin{array}{lcl}
\delta (\tilde T_{1})\; \delta (T_{1}) = \delta (\phi)\; \delta 
(T_{1}),
\end{array}\eqno(6.15)
$$
which reduces the partition function $Z$ to
$$
\begin{array}{lcl}
{\cal Z} &=& {\displaystyle \int} \; {\cal D}  [ \mu  ]\;
\;\mbox {exp} \;\Bigl ( i {\displaystyle \int} d^3 x\;
\Bigl [ |\partial_{\mu} Z |^2 - |Z^\dagger \partial_{\mu} Z|^2 + \theta\;
{\cal L}_{H} , \nonumber\\
&-& 2 \bar C_{1} C^2 + 2 \bar C_{2} C^1 + \dot C^2 \dot {\bar C_{2}}
- {\displaystyle \int}\; d^2 y \; \bar  C_{3} (x)
\sigma (x, y) C^3 (y) \Bigr ] \Bigr ), \nonumber\\
{\cal D}  [ \mu  ] &=&
{\cal D} \bigl ( z, z^*, \bar C_{2}, C^2, \bar C_{3}, C^3, \bar C_{1}, 
C^1 \bigr )\; \delta (|Z|^2 - 1)\; \delta (\chi^3).
\end{array}\eqno(6.16)
$$
Once again using the equations of motion w.r.t. $\bar C_{1}, C^1$ and 
taking the integration over $C^3$ and $\bar C_{3}$, all but the last two ghost
contributions are removed and we end up with
$$
\begin{array}{lcl}
{\cal Z} &=& {\displaystyle \int} \; {\cal D}  [ \mu  ]
\; exp \Bigl ( i {\displaystyle \int}\; d^3 x {\cal L} \Bigr ),
\nonumber\\
{\cal D}  [ \mu  ] &=&
{\cal D} \bigl ( z, z^* \bigr )\; \delta (\chi^3)\; \mbox {det} || 
\sigma ||.
\end{array}\eqno(6.17)
$$
Notice that the following choice for $\chi^3$ \cite{bbg}, mentioned in 
section II,
$$
\begin{array}{lcl}
\chi^{3} = z_{1} - z^*_{1},
\end{array}
$$
is consistent with all the restrictions imposed on $\chi^{3}$ so far and 
this choice yields
$$
\begin{array}{lcl}
\sigma (x, y) = (z_{1} + z_{1}^*)\; \delta (x - y).
\end{array}
$$
The unitary gauge is introduced to ensure the consistency of the scheme. 
More interesting gauge choices, such as the Coulomb gauge will be studied in 
a future work.\\

\noindent
{\bf 7 Conclusions}\\

\noindent
The primary motivation of this work was to construct a consistent
quantum theory of the $CP^1$ model coupled to the Hopf term, so
that eventually we can analyse the quantum effects induced by the
Hopf term. In particular, the possibility of quantum corrections to the energy
expectation value and absence of any quantum correction to fractional
spin are demonstrated. 

Due to the nonlinearity present in the original model associated
with field dependent Dirac brackets, the canonical quantization
could not be carried out because of the severe operator ordering
ambiguities. We have bypassed this problem by using Batalin-Tyutin
scheme, where the phase space is enlarged by incorporating additional
fields in such a manner that the symplectic structure in the
extended phase space is given by the usual Poisson brackets. In this
regard, we follow \cite {kor} , whereby the structures of the Hamiltonian
and the constraints (which are all first class now)
remain unaffected when written in terms of the improved variables,
thereby ensuring the existence of solitonic configurations.
The complete structure of the first class theory in the extended
space has been provided.

Using the above mentioned extended Hamiltonian, we have computed
the quantum correction to the expectation value of the energy
density, which stems from operator orderings of variables occurring
in the Hamiltonian.  Interestingly, we have found that
the Hopf term, although a topological term, can have an ${\cal O}(\hbar)$ 
finite contribution to the expectation value of the energy density at the 
quantum level in the {\it nontrivial topological sector}.
This is apart from ${\cal O}(\hbar^2)$ divergent contribution coming
from non-Hopf term in the energy expectation value.

Here we would like to emphasize the following points. Note that the canonical
quantisation of a field theoretical model is usually marred with operator 
ordering ambiguities as we mentioned earlier and these ambiguities can arise
mainly from the following two possible situations. Firstly, if the symplectic
structure given by the DBs are field dependent in a complicated manner, then
these DBs cannot be elevated to the quantum commutators consistently. This
was explicitly demonstrated in [12] in the context of this model before the
Batalin-Tyutin extension was made. But this BT extension really gets one out
of the problem as we have seen. The second problem, which persists here, even
after BT extension of this model is the non-unique hermitian expression of
any observable and Hamiltonian in particular. In equivalent operator orderings
result in inequivalent quantum theories. However, once a particular form of
this quantum Hamiltonian is chosen, it is a matter of straightforward algebra
to isolate the quantum corrections by repeated application of the simple and 
canonical  form of the commutation relations (5.3). This model, therefore,
provides a non-trivial field theoretic example involving non-linearities, where
the power of this Batalin-Tyutin quantization can be demonstrated. We should
also point out that if the BT fields are set to zero, one recovers the original
$CP^{1}$ model coupled to the Hopf term along with their complicated 
second-class constraints. This is the {\it strong} version of the unitary
gauge condition. In the quantum analysis, however, one only sets them to
{\it weakly} zero, i.e., physical states are taken to be annihilated by these
BT field operators (5.7), (5.8).

We expected a similar quantum correction in the angular momentum, giving
rise to fractional spin. However, the operator ordering does not yield any
quantum correction. It was already shown in \cite{cm} that Hopf term does
not contribute to the fractional spin at the classical level. In Section
2 of this paper, we have shown that the picture remains the same even at
the level of collective coordinate quantization. Finally, quantizing the model
utilizing the Batalin-Tyutin scheme, where there is no operator ordering 
problems anymore, we find the same picture persisting at this level too, i.e.,
no fractional spin is induced at ${\cal O} (\hbar)$ level due to quantum
effects. This result is, therefore, different from Wilczek and Zee 
\cite{wz} and furnishes another example of inequivalent quantization.  

Finally, we perform the conventional BRST quantisation of the extended model
where all the constraints have become first class. The partition function has
also been computed in the extended scheme whereby the equivalence with the
original model can also be established.


\newpage
\baselineskip = 12pt
\begin{thebibliography}{99}
\bibitem {dir} P. A. M. Dirac,  {\it Lectures on Quantum Mechanics},
       (Yeshiva University Press, New York, 1964).
\bibitem {bfv} I. A. Batalin and E. S. Fradkin {\it in}: {\it Group 
        Theoretical Methods in Physics}, Vol. II (Moscow, 1980);
        I. A. Batalin and G. A. Vilkovisky, Phys. Lett. 69 B, 309 (1977);
        M. Henneaux, Phys. Rep. 126, 1 (1985);
        M. Henneaux  and C. Teitelboim,  {\it Quantization of Gauge Systems}
       (Princeton University Press, Princeton, 1992).
\bibitem {bt} I. A. Batalin and I. V. Tyutin, Int. J. Mod. Phys. A {\bf 6}, 
              3255 (1991).
\bibitem {bbg} N. Banerjee, R. Banerjee and S. Ghosh, Phys. Rev. D {\bf 49}, 
        1996 (1994); Nucl. Phys. {\bf B 417}, 257 (1994).
\bibitem {kor} S. -T. Hong, W. T. Kim and Y. -J. Park, 
             Phys. Rev.  {\bf 60}, 125005 (1999).
\bibitem {sol} 
             R. Rajaraman,
             {\it Solitons and Instantons}, (Elsevier, North Holland, 1982);
             G. Adkins, C. Nappi and E. Witten, Nucl. Phys. 
             {\bf B228} 552 (1983); 
             A. P. Balachandran {\it in}: High Energy Physics, 
             Proceedings of Yale
             Theoretical Advanced Study Institute, eds. M. J. Bowick 
             and F. Gursey (World Scientific, Singapore, 1985).
\bibitem {kor1} S. -T. Hong, Y. -J. Park, K. Kubodera and F. Myhrer,
          {\it Improved Dirac quantization of $CP^1$ model}: 
          hep-th/0006227.
\bibitem {fradkin} E. Fradkin,{\it Field Theories in Condensed Matter Systems}
         (Addition-Wesley, 1991).
\bibitem {gov} B. Chakraborty and T. R. Govindarajan, 
         Mod. Phys. Lett. A {\bf 12}, 619 (1997).
\bibitem {wz} F. Wilczek and A. Zee, Phys. Rev. Lett. {\bf 51} 2250 (1983).
\bibitem {bkw} M. Bowick, D. Karabali and L. C. R. Wijewardhana,;
          Nucl. Phys. {\bf B 271}, 417 (1986).
\bibitem {cm} B. Chakraborty and A. S. Majumdar, Int. J. Mod. Phys. A 
         {\bf 14}, 1561 (1999);
         Phys. Rev. D {\bf 58}, 125024 (1998); Acta. Phys. Pol. B 
         {\bf 30}, 247 (1999).
\bibitem {no} W. Oliveira and J. A. Neto, Int. J. Mod. Phys. A {\bf 12}, 4895 
         (1997).
\bibitem {wuz} Y. Wu and A. Zee, Phys. Lett. B {\bf 147} 325 (1984).
\bibitem {bc} R. Banerjee and B. Chakraborty, Nucl. Phys. 
         {\bf B 449}, 317 (1995).
\bibitem {forte} S. Forte, Rev. Mod. Phys. {\bf 64}, 193 (1992).
\bibitem {lee} See, e.g., T. D. Lee, {\it Particle Physics and Introduction
          to Field Theory}, ( Harwood, N. Y., 1981).
\end{thebibliography}
\end{document}